\documentclass{jfm}
\usepackage{graphicx}
\usepackage{natbib}
\usepackage{amsmath}
 \usepackage{float}
\usepackage{bm}
\usepackage[utf8]{inputenc}
\usepackage{enumitem}
\usepackage{subcaption}
\usepackage{xcolor}
\usepackage{placeins}

\title{Reduced kinetic model of polyatomic gases}
\author{Praveen Kumar Kolluru, 
        Mohammad Atif \and
        Santosh Ansumali \corresp{\email{ansumali@jncasr.ac.in}}}
\affiliation{Jawaharlal Nehru Centre for Advanced Scientific Research,
             Jakkur, Bangalore  560064, India}

\begin{document}
\maketitle
\begin{abstract}
Kinetic models of polyatomic gas typically account for the internal degrees of freedom at the level of the two-particle distribution function. However, close to the hydrodynamic limit, the internal (rotational) degrees of freedom tend to be well represented just by rotational kinetic energy density. We account for the rotational energy by augmenting the Ellipsoidal-statistical BGK (ES--BGK) model, an extension of the Bhatnagar--Gross--Krook (BGK) model, at the level of the single-particle distribution function with an advection-diffusion-relaxation equation for the rotational energy.
 This reduced model respects the $H$ theorem and recovers the
compressible hydrodynamics for polyatomic gases as its macroscopic limit. As required for a polyatomic gas model, this extension of the ES--BGK model has not only correct specific heat ratio but also allows for three independent tunable transport coefficients: thermal conductivity,  shear viscosity, and bulk viscosity. We illustrate the effectiveness of the model via a lattice Boltzmann method implementation.
\end{abstract}

\section{Introduction}
The dynamics of a dilute monoatomic gas in terms of the single-particle distribution function  is described by the Boltzmann equation   \citep{chapman1970mathematical, cercignani1988boltzmann}. 
Unlike the continuum  Navier--Stokes--Fourier hydrodynamics equation, the Boltzmann equation is a valid description even at highly non-equilibrium states  \citep{mott1951solution,liepmann1962structure,cercignani1988boltzmann}, encountered   in the presence of strong shock waves at high Mach number (ratio of flow speed to sound speed) and in a highly rarefied flow characterized by a large Knudsen number (ratio of the mean free path to characteristic length scale)  \citep{oh1997computations,struchtrup2004stable,
ansumali2007hydrodynamics}. 
However,  any   analysis of the integro-differential Boltzmann equation is a formidable task even for the simplest  problems. Thus, one often models the Boltzmann dynamics via a simplified collision term that converts the evolution equation to a partial differential equation  \citep{bhatnagar1954model,lebowitz1960nonequilibrium, holway1966new, 
 shakhov1968generalization, gorban1994general, levermore1996moment, 
 andries2000gaussian, ansumali2007quasi,lebowitz1960nonequilibrium, 
singh2015fokker,agrawal_singh_ansumali_2020}.  
An important example  is the  BGK model (Bhatnagar et al. 1954),
which states that the relaxation of the distribution function towards the Maxwell--Boltzmann (MB) form happens in a time scale corresponding to   the mean free time $\tau$ with the assumption that every moment of the distribution function relaxes at the same rate. The BGK model is quite successful in replicating qualitative features of the Boltzmann dynamics (collisional invariants, the zero of the collision, H-theorem, conservation laws, etc).
However,  the BGK model predicts the Prandtl number of the fluid to be unity, while the value predicted by the Boltzmann equation for monoatomic gas is $2/3$. Thus, several other variations of the collision model such as ES--BGK model  \citep{holway1966new,andries2000gaussian}, the quasi-equilibrium models  \citep{gorban1994general,levermore1996moment}, the Shakhov model  \citep{shakhov1968generalization}, and the Fokker-Plank model  \citep{singh2015fokker, singh2016gaseous} are used as kinetic models with tunable Prandtl number. The ES--BGK model  \citep{holway1966new,andries2000gaussian} is   an elegant  but simple improvement over the BGK model. This model assumes that the distribution function relaxes to an anisotropic Gaussian distribution within mean free time $\tau$. The anisotropic Gaussian in itself evolves towards the MB distribution with a second time scale. 
The presence of a second time scale as free parameter  ensures that the time scales related to momentum and thermal diffusivity are independent and thus permits one to vary the Prandtl number in the range of $2/3$ to $\infty$.



Despite their success, the Boltzmann collision kernel and its aforementioned simplifications are limited to monoatomic gases as they do not account for the internal molecular structure. 
However, many real gases such as nitrogen, oxygen, or methane are polyatomic.
At the macroscopic level, the internal molecular structure predominantly manifests in terms of modified specific heat ratio $\gamma$ and bulk viscosity $\eta_{\rm B}$, which is crucial for a number of aerodynamic and turbomachinery engineering applications  \citep{von2008effect,wu2015kinetic}. 
The specific heat ratio predicted by the Boltzmann equation is that of a monoatomic gas ($\gamma=5/3$), whereas that of a diatomic gas is $7/5$.

Two-particle kinetic theory as an extension of the Boltzmann equation  as expected  correctly predicts the specific heat ratio for polyatomic gases along with heat conductivity and the bulk viscosity  \citep{wang1951transport,wu2015kinetic,chapman1970mathematical}.
However, it is often not feasible to do any  analysis on the Boltzmann-type equation for polyatomic gases. 
Therefore, several simplifications to model polyatomic gases have also been proposed. 
They essentially incorporate the rotational kinetic energy by decomposing the two-particle 
distribution function into two independent single-particle distribution 
functions  \citep{morse1964kinetic, andries2000gaussian, tsutahara2008new, 
kataoka2004lattice, watari2007finite, nie2008thermal, 
larina2010kinetic, wu2015kinetic, wang2017unified, 
bernard2019bgk}. 
Furthermore, a thermodynamic framework and extensions thereof were developed for modelling highly nonequilibrium phenomena in dense and rarefied polyatomic gases where the Navier--Stokes--Fourier theory is no longer valid  \citep{muller2013rational, ruggeri2015rational, 
arima2012extended}. 
A few BGK like models have also been proposed for polyatomic gases which accept the Prandtl number as a tunable parameter  \citep{andries2000gaussian,brull2009ellipsoidal}. 

Hydrodynamic simulations for a realistic system require the development of reduced-order models to account for rotational degrees of freedom ideally without increasing the phase-space dimensionality. 
Indeed, the standard approach is to demonstrate that the two-particle distribution function describing the translational and rotational degree of freedom can be approximated by considering two single-particle distribution functions (one each for the translational and rotational degree of freedom) whose dynamics are coupled to each other  \citep{andries2000gaussian}. However, recently it was pointed out that a simplified description in terms of single-particle distribution function for the translational degree of freedom and a scalar field variable for rotational kinetic energy is sufficient for modelling the change in specific heat ratio for a dilute diatomic gas in the hydrodynamic limit  \citep{kolluru2020extended}. This model supplemented the standard Boltzmann BGK equation with an advection-relaxation equation for the evolution of rotational energy. It preserved the correct conservation laws for diatomic gases in the hydrodynamic limit and satisfied the $H$ theorem. However, the model was restricted to diatomic gases and a Prandtl number $7/5$, limiting its application for heat transfer problems.

We propose a kinetic model of polyatomic gases  to tune Prandtl number, specific heat ratio, and bulk viscosity in a physically transparent fashion.
 To do so, we write a new  collision kernel which is a linear combination of the ES--BGK and BGK kernels that are locally relaxing to different temperatures at different timescales. The ratio of the two relaxation timescales  is used to tune the Prandtl number. We couple the evolution of the single-particle distribution function (with this modified collision kernel) via  an advection-diffusion-relaxation equation for the rotational energy.  The rotational contribution to the internal energy alters the specific heat ratio to that of a polyatomic gas and allows to model bulk viscosity contribution arising out of the rotational degree of freedom. 
Such an extension of the ES--BGK model indeed reproduces the hydrodynamic behaviour of a polyatomic gas and also has a valid $H$ theorem.   These minimal extensions of the ES--BGK model of monoatomic gas are constructed at a single-particle level for polyatomic gases and are phenomenological by construction. 
It is commensurate with the top-down modelling approach as developed in the context of the lattice Boltzmann models and aim to be analytically and numerically tractable  \citep{succi2001lattice,ansumali2007hydrodynamics,atif2017}. The present model which requires only the solution of an advection-diffusion-relaxation equation along with the Boltzmann ES--BGK equation adds only a  minor complexity over analogous monoatomic gas ES--BGK model and can be implemented in the mesoscale framework such as lattice Boltzmann (LB) method quite easily. This approach is distinctly different and is more detailed than the existing approach in the LB models where the effect of rotational degree of freedom is further coarse-grained and the correction needed to model specific heat ratio is directly added as a force term in the BGK collision model    \citep{kataoka2004lattice,nie2008thermal,chen2010multiple,huang2020lattice}. In contrast, this model of polyatomic gas  enlarges the set of microscopic degrees of freedom and  models dynamics of rotational energy in an explicit manner.

The manuscript is organized as follows: A brief kinetic description of
monoatomic and polyatomic gases is given in Sections \ref{sec:kinMonotomic} and 
\ref{sec:kinDiatomic} respectively. In Section \ref{sec:ES--BGK-diatomic} we 
propose an extension to the ES--BGK model for polyatomic gases. The lattice Boltzmann formulation is described in 
Section \ref{sec:numMtd}. The proposed model is
numerically validated in Section \ref{sec:validation}.

\section{Kinetic description of a monoatomic gas} \label{sec:kinMonotomic}

The dynamics of dilute monoatomic gases is well-described by the Boltzmann 
equation in terms of  the evolution of the  single-particle distribution 
function $f$, where $f({\bm x}, {\bm c}, t) \, d{\bm x} \, d {\bm c}$ is the 
probability of finding a particle within $({\bm x}, {\bm x} 
+ d{\bm x})$, possessing a velocity in the range $( {\bm c},  {\bm c} + d {\bf 
c})$ at a time $t$. The hydrodynamic variables are density $\rho({\bm x}, t)$, 
velocity ${\bm u}({\bm x}, t)$ and total energy $E({\bm x}, t)=(\rho u^2 + 3 
\rho k_B T /m)/2$. Here onwards, we use a scaled temperature $\theta$ 
defined in terms of Boltzmann constant $k_B$ and mass of the 
particle $m$ as $\theta = k_B T /m$.   The thermodynamic pressure $p$ and the 
scaled temperature $\theta$ are related via the ideal gas equation of state as 
$p = \rho \theta$. These hydrodynamic variables are computed as the moments of 
the single-particle distribution function   
\begin{align}
\begin{split}
\{ \rho, \rho {\bm u}, E \} = \left\langle \left\{ 1, {\bm c}, \frac{c^2}{2} \right\}, f \right\rangle,
\end{split}
\label{hydrodynamicMoments}
\end{align}
where  we define the averaging operator 
$\langle \phi_1({\bm c}), \phi_2({\bm c}) \rangle 
 = \int_{-\infty}^{\infty} \phi_1({\bm c}) \phi_2({\bm c})d{\bm c}$.
In the comoving reference frame with fluctuating velocity ${\bm{\xi}} = {\bm c}$ - ${\bm u}$, the stress tensor $\sigma_{\alpha \beta}$ is 
traceless part of the flux of the     momentum tensor $\Theta_{ij}\equiv = \sigma_{\alpha \beta}+\rho \theta \delta_{\alpha \beta}$ and the heat flux $q_{\alpha}$ 
is the flux of  the energy. Thus, 
\begin{equation}
 \Theta_{ij}\equiv \left< \xi_i \xi_j, f\right>,\; \; \sigma_{\alpha \beta} = \left< \overline{\xi_{\alpha} \xi_{\beta}}, f\right>, 
\quad q_{\alpha} = \left<\xi_{\alpha} \frac{\xi^2}{2}, 
f\right>,\quad 
\end{equation}
where the symmetrized traceless part $\overline{A_{\alpha \beta}}$ for any 
second-order tensor $A_{\alpha \beta}$  is $\overline{A_{\alpha \beta}} =\left( 
A_{\alpha \beta} + A_{ \beta \alpha}  - 2 A_{\gamma \gamma} \delta_{\alpha 
\beta}/3 \right)/2$.  The stress tensor and heat flux tensor are related to the 
pressure tensor $P_{\alpha \beta} = \left<c_{\alpha} c_{\beta}, f\right>$ and 
energy flux  $\left\langle c^2 c_\alpha/2,f \right\rangle$ respectively as 
\begin{align}
\begin{split}
 \sigma_{\alpha \beta} &= P_{\alpha \beta} - \rho u_{\alpha} u_{\beta} - 
\rho \theta \delta_{\alpha \beta}, \qquad
 q_{\alpha} = \left\langle \frac{c^2 c_\alpha}{2},f \right\rangle - u_\alpha 
\left( E + \rho \theta \right) - u_\beta \sigma_{\alpha \beta}.
\end{split}
\end{align}
We also  define 
  third-order moment $Q_{\alpha \beta \gamma}$, with  the traceless part  $\overline{Q_{\alpha \beta \gamma}}$ and the  fourth-order moments divided into contracted part $R_{\alpha \beta}$ and the trace $R$  
\begin{equation} 
 Q_{\alpha \beta \gamma}= \left<\xi_{\alpha} 
\xi_{\beta}  \xi_{\gamma}, f\right>, \quad  \overline{Q_{\alpha \beta \gamma}} = Q_{\alpha 
\beta \gamma} - \frac{2}{5} \left( q_{\alpha} \delta_{\beta \gamma} 
+ q_{\beta} \delta_{ \alpha \gamma} + q_{\gamma} \delta_{\alpha \beta} 
\right),
\end{equation} 
\begin{equation}
R_{\alpha \beta} = 
\left< \xi^2 \xi_{\alpha} \xi_{\beta}, f\right>,\quad R = \left< \xi^4, f \right>.
\end{equation}

In the dilute gas limit, the time evolution of the distribution function is  a sequence of free-flight and binary collisions well described by the  Boltzmann equation 
\begin{align}
\begin{split}
 \partial_t f  + c_\alpha \partial_\alpha f = \Omega(f,f),
\end{split}
\label{eq:BoltzmannEq}
\end{align}
where the collision kernel $\Omega(f,f)$ models the binary collisions between particles under the assumptions of molecular chaos  \citep{chapman1970mathematical,cercignani1988boltzmann,mcquarrie2000statistical}. 
The nonlinear integro-differential Boltzmann collision kernel is often replaced by simpler models that should recover the following essential features  \citep{cercignani1988boltzmann}:

\begin{enumerate} [label=(\alph*)]
 \item \textbf{ Conservation Laws:}  As the mass, momentum, and energy density of the 
particles are conserved during the elastic collision, any collision model must 
satisfy 
\begin{equation} 
\langle \Omega(f,f), \{ 1,{\bm c},c^2 \} \rangle  = \{ 0,\mathbf{0},0 \}.
\end{equation}
Thus, the macroscopic conservation laws obtained by taking appropriate 
moments (with respect to $\{ 1,{\bm c},c^2 \}$) of the Boltzmann equation   
\begin{align}
\begin{split}
   &\partial_t \rho + \partial_\alpha \left( \rho u_\alpha 
\right) = 0, \\
   &\partial_t \left(\rho u_\alpha \right) + \partial_\beta \left( \rho 
u_\alpha u_\beta + \rho \theta \delta_{\alpha \beta} + \sigma_{\alpha 
\beta}\right) = 0, 
\\&\partial_t E + 
     \partial_\alpha \left[ \left( E + p \right) 
 u_\alpha + q_\alpha + \sigma_{\alpha \beta} u_\beta \right] =  0,
\end{split}
\end{align}
 are the same as those of the compressible hydrodynamics.

\item \textbf{ Zero of collision:}  The   collision term is zero if and only if  
the distribution function attains Maxwell--Boltzmann form, i.e.,
\begin{equation}
 \Omega(f,f) = 0 \iff f = f^{\rm{MB}}, 
\end{equation}
where  the  Maxwell--Boltzmann distribution $f^{\rm MB}$  is
\begin{equation}
  f^{\rm MB}(\rho(f),{\bm u}(f),\theta(f))  = \rho\,\left(\frac{1}{2 \pi 
\theta} 
\right)^{3/2}  \exp{\left(-\frac{(c-u)^2}{2\, \theta}\right) },
 \label{MaxwellBoltzmann}
\end{equation}
which ensures that the dynamics is consistent with  the  equilibrium 
thermodynamics.

\item \textbf{ The H Theorem: } The Boltzmann equation generalizes the second law of 
thermodynamics to nonequilibrium situations. 
Boltzmann defined a nonequilibrium generalization of the entropy known as the $H$ function
 \citep{cercignani1988boltzmann}
\begin{equation}
\label{h--func}
H[f] = \int (f \ln f - f) dc.
\end{equation}
At equilibrium, the thermodynamic entropy is calculated as    \citep{cercignani1988boltzmann}
\begin{equation}
S^{\rm eq}=-k_{\rm B} H[f^{\rm MB}] =
- \rho k_B  \left[\ln{ \frac{\rho}{\left(2 \pi \theta \right)^{3/2} } - \frac{5}{2}}
\right]. 
\end{equation}

The evolution of this $H$ function is  
\begin{equation}
\partial_t H + \partial_\alpha J_\alpha^{H} = 
\underbrace{\left\langle \Omega(f,f), \ln f \right\rangle}_{\Sigma} \leq 0,
\end{equation}
where  $J_\alpha^{H}$ is the entropy flux and $\Sigma$ is the negative of the entropy generation. Boltzmann demonstrated that the entropy generation is positive,
hence, the $H$ function is nonincreasing in time  \citep{cercignani1988boltzmann}. 
Thus, any consistent approximation for the Boltzmann collision kernel should 
also satisfy the same condition.
\end{enumerate}

 We briefly describe the two most widely used models, the BGK 
collision model and  the ES--BGK model. The Bhatnagar--Gross--Krook (BGK) model, perhaps the simplest and most widely-used model of the Boltzmann collision kernel  models the  collision as a relaxation of the distribution function towards the equilibrium $f^{\rm{MB}}$ 
  \citep{bhatnagar1954model} 
 as
\begin{equation}
 \Omega_{\rm{BGK}} = \frac{1}{\tau} \left(f^{\rm{MB}}(\rho,{\bm u}, \theta) 
- f \right).
\label{BGK}
\end{equation}
This assumes that the process occurs at a single time scale 
$\tau$ corresponding to the mean free time.  It is trivial to see that this 
model has the same collisional invariants as the Boltzmann kernel, hence, recovers the same conservation laws  \citep{bhatnagar1954model}.
The entropy production $\Sigma^{\rm BGK}$ due to the BGK model is written as
\begin{align}
\begin{split}
 \Sigma^{\rm{BGK}} = 
  \left\langle \Omega_{\rm{BGK}}, \ln f \right\rangle = 
\frac{1}{\tau} \int \left(f^{\rm{MB}}(\rho,{\bm u}, 
\theta) - f \right)  \ln \left( \frac{f}{ f^{\rm MB} (\rho,{\bm u}, 
\theta)} \right) d{\bm c} \leq 0.
\end{split}
\end{align}
Thus, like the Boltzmann equation, the BGK model also satisfies the $H$ theorem.  By taking 
appropriate moments of the Boltzmann BGK equation, we can also see that the 
evolution of the stress and the heat flux are  \citep{liboff2003kinetic}
 
\begin{align}
  \begin{split}
 \partial_t \sigma_{\alpha \beta}
 +  \partial_\gamma \left( \sigma_{\alpha \beta} u_\gamma \right)
 +  \partial_\gamma \overline{Q_{\alpha \beta \gamma}} 
 + 2 \overline{\sigma_{\gamma \beta} \partial_\gamma u_\alpha}
 + 2 p \overline{\partial_\beta u_\alpha}
 + \frac{4}{5} \overline{\partial_\beta q_\alpha} 
 = -\frac{1}{\tau} \sigma_{\alpha \beta}, 
\\
 \partial_t {q}_{\alpha} + \frac{1}{2} \partial_{\beta} \left( 
\overline{R_{\alpha \beta}} 
 + \frac{1}{3}R\,\delta_{\alpha\beta} \right)+ \overline{ Q_{\alpha \beta 
\gamma} } \partial_{\gamma} 
u_{\beta} + \partial_{\beta} \left( q_{\alpha} u_{\beta} \right) 
 + \frac{7}{5}  q_{\beta} \partial_{\beta} u_{\alpha} \\
+ \frac{2}{5}  
q_{\alpha} \partial_{\beta} u_{\beta}  + \frac{2}{5}  q_{\beta} 
\partial_{\alpha} u_{\beta}  -\frac{5}{2} \frac{p}{\rho} \partial_{\alpha}p
- \frac{5}{2} \frac{p}{\rho} \partial_{\beta} \sigma_{\alpha \beta}
 - \frac{\sigma_{\alpha \beta}}{\rho} \partial_{\beta}p
 - \frac{\sigma_{\alpha 
\beta}}{\rho} \partial_{\eta} \sigma_{\beta \eta}  =  
-\frac{{ q}_{\alpha}}{\tau }.
\end{split}
\end{align}
The form of relaxation dynamics shows that the time-scales for the 
momentum diffusivity and thermal diffusivity are equal for the BGK model. These equations show that,  like any equation of Boltzmann
 type, the dynamics of $M^{th}$ order moment involves $(M+1)^{th}$ moment and thus form an infinite order moment chain.  The hydrodynamic limit is typically analysed via Chapman--Enskog expansion which allows evaluating
  the dynamic viscosity $\mu$  and thermal conductivity $\kappa$ for 
the model  with the specific heat $C_p$ for a monoatomic ideal gas as $5/2$ is   \citep{chapman1970mathematical}
\begin{equation}
 \mu = p \tau ,\quad \kappa = \frac{5}{2} p \tau \quad \implies {\rm Pr} = \frac{\mu C_p}{k} = 1.
\end{equation}
 Despite this defect of   ${\rm Pr}=1$, the BGK model is extremely successful both as a numerical and an analytical tool for analysis.
The ES--BGK model  \citep{holway1965kinetic} also describes the collision as simple relaxation process  but unlike the BGK model, it overcomes the restriction on the Prandtl number. The extra ingredient for ES--BGK model is quasi-equilibrium form of distribution  derived by minimizing the H--function (Eq.\eqref{h--func}) under the constraints of an additional condition of fixed stresses, which implies absolute minimization of 
 
\begin{equation}
\Xi[f]=  \int d{\bf c} \left[(f \ln f - f)  +\alpha   f +\beta_j c_j f + 
\gamma_{i j} c_i c_j f\right].
\end{equation}
The solution to this minimization problem is an anisotropic Gaussian 
\begin{equation}
 f^{\rm Quasi}\left(\rho, \bm{u}, \Theta_{ij} \right) =  \frac{\rho}{\sqrt{\rm 
{det}[2 \pi \Theta_{ij}]}}  \exp\left({-\frac{1}{2} \xi_i \Theta_{ij}^{-1}
\xi_j} \right).
\end{equation}
  Like the MB distribution, this has the same conserved moments as that of the single-particle distribution function  but  this also treats  $\Theta_{ij}(f)$ as a conserved variable  \citep{kogan1969rarefied} i.e.,
\begin{equation}
\{\rho(f^{\rm Quasi}) , u_{\alpha}(f^{\rm Quasi}), \theta(f^{\rm Quasi}),\Theta_{ij}(f^{\rm Quasi}) \} 
= \{\rho(f) , u_\alpha(f), \theta(f),\Theta_{ij}(f) \}.
\end{equation}
 On this quasi-equilibrium manifold with stress as variable when the $H$ is minimum we have
\begin{equation}
\label{HQuasi}
 H[f^{\rm Quasi}\left(\rho, \bm{u}, \Theta_{ij} \right)] =
 \rho \, \ln{\left(\frac{\rho}{\sqrt{\rm 
{det}[2 \pi \Theta_{ij}]}} \right)}  -\frac{5}{2} \rho.
\end{equation}

The ES--BGK model uses the anisotropic Gaussian distribution  $f^{\rm ES}\equiv f^{\rm Quasi}\left(\rho, \bm{u}, \lambda_{ij} \right)$ 
\begin{equation}
 f^{\rm ES}\left(\rho, \bm{u}, \lambda_{ij} \right) =  \frac{\rho}{\sqrt{\rm 
{det}[2 \pi \lambda_{ij}]}}  \exp\left({-\frac{1}{2} \xi_i \lambda^{-1}_{ij} 
\xi_j} \right),
\end{equation}
where instead of pressure tensor a positive definite matrix
$ \lambda_{ij} $ 
\begin{equation}
 \lambda_{ij} = (1-b) \theta \delta_{ij} + b \Theta_{ij}\equiv
 \theta \delta_{ij} + b \sigma_{ij},
 \label{lambdaij}
\end{equation}
is used with  $-1/2 \leq b \leq 1$ as a free parameter, the range of $b$ is dictated by 
the positive definiteness  of $\lambda^{-1}_{ij}$. For $0 \leq b \leq 1$, 
$\lambda_{ij}$ is trivially positive as it is a convex combination of two 
positive definite matrices. The non-trivial range $-1/2\leq b<0$ is better understood from the eigenvalue analysis  of $\lambda_{ij}$   \citep{andries2000gaussian}. Let the 
eigenvalues of $\lambda_{ij} $ be $\Lambda_i$ and that of positive definite 
matrix $\Theta_{ij}$ be $\phi_i$ and  thus  
$\phi_1+\phi_2+\phi_3=\Lambda_1+\Lambda_2+\Lambda_3=3 \theta$.
In terms of these eigenvalues, the matrix ${\bm \lambda}$ after suitable rotation can be rewritten as
\begin{equation}
{\bm \lambda}=
 \begin{pmatrix}
  (1-b)\theta+b \phi_1 &0&0\\
  0&  (1-b)\theta+b \phi_2&0\\
  0 & 0 & (1-b)\theta+ b\phi_3
 \end{pmatrix},
 \label{diagrep}
\end{equation}
 which is a  convex combination of
diagonal matrices $\Psi_1, \Psi_2$ and $\Psi_3$ as
\begin{equation}
\label{BMForm}
{\bm \lambda} =  \left( \frac{1+2b}{3} \right) \underbrace{\begin{pmatrix}
  \phi_1 & 0 & 0 \\
  0 & \phi_2 & 0 \\
  0 & 0 & \phi_3
 \end{pmatrix} }_{\Psi_1}+
                \left( \frac{1-b}{3} \right) \underbrace{\begin{pmatrix}
  \phi_2 & 0 & 0 \\
  0 & \phi_3 & 0 \\
  0 & 0 & \phi_1
 \end{pmatrix} }_{\Psi_2} +
                \left(\frac{1-b}{3} \right) \underbrace{\begin{pmatrix}
  \phi_3 & 0 & 0 \\
  0 & \phi_1 & 0 \\
  0 & 0 & \phi_2
 \end{pmatrix} }_{\Psi_3}.
\end{equation}
  As in the diagonal representation, the non-zero components are the eigenvalues
  and one obtains the  relationship between  $\phi_i$  and $\Lambda_i$ as
\begin{align}
 \begin{split}
 \Lambda_i =  \left( \frac{1+2b}{3}\right) \phi_i +
          \left(\frac{1-b}{3} \right) \sum_{i \neq j} \phi_j.
 \end{split}
 \label{diagEigen}
\end{align}
Thus, the eigenvalues of $\lambda_{ij}$ are non-negative if the range of $b$ is restricted to 
$-1/2 \leq b \leq 1$ as we also know that  $\phi_i \geq 0$. In the ES--BGK model, the collisional relaxation is towards this anisotropic Gaussian distribution $f^{\rm ES}$
 which itself attains the form of the Maxwellian at the equilibrium.  The collision kernel in explicit form is 
\begin{equation}
 \Omega_{\rm ESBGK}= \frac{1}{\tau} 
 \left( f^{\rm ES}\left(\rho, \bm{u}, \lambda_{ij} \right) - f\right),
\label{eq:ES--BGK}
\end{equation}
where it is evident that $b=0$ corresponds to the limit of the BGK equation and $b=1$ would imply that stress is conserved. For $b \neq 1$, the model has the same set of conservation laws as the BGK equation.
 As the stress and heat flux tensors follow the relation 
\begin{align}
\begin{split}
\sigma_{\alpha \beta}( f^{\rm ES})
=  b\sigma_{\alpha \beta}( f), \quad 
\left\langle  \xi_{\alpha} \xi^2 , f^{\rm ES} \right\rangle = {0},
\end{split}
\end{align}
  thus,  the evolution equations for stress tensor and heat flux are
\begin{align}
  \begin{split}
 \partial_t \sigma_{\alpha \beta} +  \partial_\gamma  \left(u_\gamma \sigma_{\alpha 
\beta} \right) + \partial_\gamma \overline{Q_{\alpha \beta \gamma}}
+ 2 \overline{\sigma_{\gamma \beta} 
\partial_\gamma u_\alpha} + 2 p \overline{\partial_\beta u_\alpha} + \frac{4}{5}
\overline{\partial_\beta q_\alpha} = -\left(\frac{1-b}{\tau}\right) 
\sigma_{\alpha \beta},\\
 \partial_t {q}_{\alpha} + \frac{1}{2} \partial_{\beta} \left( 
\overline{R_{\alpha \beta}} 
 + \frac{1}{3}R\,\delta_{\alpha\beta} \right)+ \overline{ Q_{\alpha \beta 
\gamma} } \partial_{\gamma} 
u_{\beta} + \partial_{\beta} \left( q_{\alpha} u_{\beta} \right) 
 + \frac{7}{5}  q_{\beta} \partial_{\beta} u_{\alpha} \\
+ \frac{2}{5}  
q_{\alpha} \partial_{\beta} u_{\beta}  + \frac{2}{5}  q_{\beta} 
\partial_{\alpha} u_{\beta}  -\frac{5}{2} \frac{p}{\rho} \partial_{\alpha}p
- \frac{5}{2} \frac{p}{\rho} \partial_{\beta} \sigma_{\alpha \beta}
 - \frac{\sigma_{\alpha \beta}}{\rho} \partial_{\beta}p
 - \frac{\sigma_{\alpha 
\beta}}{\rho} \partial_{\eta} \sigma_{\beta \eta}  =  
-\frac{{ q}_{\alpha}}{\tau },
\end{split}
\end{align}
which shows that the momentum and thermal diffusivities are different in an ES--BGK model and the Prandtl number ${\rm Pr} = \mu c_p / \kappa$ is a free parameter. In particular, the Chapman-Enskog analysis of this model yields 
\begin{equation}
 \mu = \frac{p \tau}{1-b} ,\quad \kappa = \frac{5}{2} p \tau\quad \implies {\rm Pr} = \frac{1}{1-b}.
\end{equation}
  
Thus, the free parameter $b$ in the anisotropic Gaussian allows one to vary the Prandtl number from $2/3$ to infinity.  At $b=-1/2$, the Prandtl number predicted by the ES--BGK model is $2/3$, which matches with the value obtained from the Boltzmann equation, and when $b=0$, the model is equivalent to the BGK model. The thermal conductivity is fixed only by $\tau$ while the viscosity can be tuned via $b$ to obtain the required Prandtl number.

 The $H$ theorem for this model was first proved by \citet{andries2001bgk} by showing that the entropy production $\Sigma^{\rm ESBGK}$ is non-positive.
 We briefly review the proof of the $H$ theorem for the ES-BGK model. The expression for the entropy production is 
\begin{align}
\Sigma^{\rm ESBGK} =  \left\langle \Omega_{\rm ESBGK}, \ln f \right\rangle  
= \frac{1}{\tau} \left\langle\left( f^{\rm ES}\left(\rho, \bm{u}, \lambda_{ij} \right) - f\right),  \frac{\partial H }{ \partial f}  \right\rangle.
\end{align}
 
The proof   is built on the property of an arbitrary  convex function $G(x)$ that
\begin{equation}
  \frac{\partial G}{\partial x}\left(y-x \right)\leq G(y) -G(x)
\end{equation}
 using which we can write 
\begin{align}
\Sigma^{\rm ESBGK} \leq  \frac{1}{\tau} \left( H[f^{\rm ES}\left(\rho, \bm{u}, \lambda_{ij} \right)] - H \left[ f^{\rm Quasi} \left(\rho, \bm{u}, \Theta_{ij} \right) \right] \right)  
+  \frac{1}{\tau} \Delta H^{\rm Quasi},
\label{appen2}
\end{align}
 with the last term as $\Delta H^{\rm Quasi}= H [ f^{\rm ES}\left(\rho, \bm{u}, \Theta_{ij} \right) ] - H[f]$ in the above equation is non-positive as $f^{\rm Quasi}\left(\rho, \bm{u}, \Theta_{ij} \right)$ is by construction the minima of $H$. To prove that 
$H[f^{\rm ES}\left(\rho, \bm{u}, \lambda_{ij} \right)] - H \left[ f^{\rm Quasi} \left(\rho, \bm{u}, \Theta_{ij} \right) \right] \leq 0$,
 using Eq.\eqref{HQuasi}  we have
\begin{align}
  H[f^{\rm ES}\left(\rho, \bm{u}, \lambda_{ij} \right)] - H \left[ f^{\rm Quasi} \left(\rho, \bm{u}, \Theta_{ij} \right) \right] = \frac{1}{2} \rho \ln \left( \frac{{\rm det} [\Theta_{ij}] }{{\rm det} [\lambda_{ij}] } \right).
\end{align}
Starting from the  Brunn-Minkowsky inequality 
\begin{equation}
 {\rm det}[aA + (1-a)B] \geq \left( {\rm det}[A] \right)^a \left( {\rm det}[B] \right)^{1-a},
\end{equation} 
relating  the determinants of two positive matrices $A$ and $B$ and their 
convex combinations, we can show that ${\rm det}[\lambda_{ij}] \geq {\rm det} [\Theta_{ij}]$   \citep{horn2012matrix}.
This inequality along with Eq.\eqref{BMForm} allows us to write
\begin{equation}
 {\rm det}[\lambda_{ij}] \geq  \left({\rm det} [\Psi_1] \right)^{\frac{1+2b}{3}}
                               \left({\rm det} [\Psi_2] \right)^{\frac{1-b}{3} }
                               \left({\rm det} [\Psi_3] \right)^{\frac{1-b}{3} }.
\end{equation}
However, from the definitions of $\Psi_1, \Psi_2$, and $\Psi_3$ one can see that
\begin{equation}
{\rm det} [\Psi_1]  = {\rm det} [\Psi_2] = {\rm det} [\Psi_3] = \phi_1 \phi_2 \phi_3
= {\rm det} [\Theta_{ij}].
\end{equation}
  Hence, the total entropy production is non-positive, i.e., $\Sigma^{\rm ESBGK} \leq 0$.

\section{Kinetic description of a polyatomic gas} \label{sec:kinDiatomic}

The rotational degrees of freedom of a polyatomic gas manifest themselves at 
the continuum level in terms of change in specific heat ratio $\gamma$ and a non-zero bulk viscosity due to interaction among the translational component $E_{\rm T} = \rho u^2/2 + 3\rho \theta_T/2$  and rotational component $E_{\rm R}$ of energy.
Thus, the  rotational degrees of freedom need to be explicitly accounted for in any microscopic or kinetic description. Indeed, typically the kinetic descriptions are in terms of a two-particle distribution function $F({\bm x}, {\bm c}, t, I)$ which defines the probability of finding a molecule with a position in the range 
$({\bm x}, {\bm x}+ d{\bm x})$ possessing a velocity in 
the range $(\bm{c}, \bm{c}+ \bm{dc})$ with an internal energy $(I, I+ dI)$ due to 
the additional degrees of freedom  \citep{morse1964kinetic, rykov1975model, kuvsvcer1989model}. For a polyatomic gas with $\delta$ additional rotational degrees of freedom, the moments of this distribution 
function  give the density, momentum, and total energy (with $\delta=0$ corresponding to a monoatomic gas)
\begin{align}
\begin{split}
 \{ \rho, \rho {\bm u},  E_{T}  +  E_{R} \} = \left\langle\left\langle \left\{ 1, {\bm c}, \frac{c^2}{2} + I^{2/\delta} \right\}, F \right\rangle\right\rangle,
\end{split}
\label{FMoments}
\end{align}
like its monoatomic counterpart and the operator $\left< \left< , \right> \right>$ is defined as
\begin{equation}
  \left \langle \left\langle \phi_1({\bm c}, I), \phi_2({\bm c}, I) 
\right \rangle \right \rangle = \int \int \phi_1({\bm c}, I) \phi_2( {\bm c}, 
I) d{\bm c} dI.
\end{equation}

For the reduced-order modeling, the distribution function $F({\bm x}, {\bm c}, t, 
I)$ is often split into two coupled distribution functions $f_1({\bm x}, {\bm c}, t)$  and $f_2({\bm x}, {\bm c}, t)$ defined as
\begin{align}
\begin{split}
 f_1({\bm x}, {\bm c}, t) = \int F({\bm x}, {\bm c}, I, t) dI, \quad \quad
 f_2({\bm x}, {\bm c}, t) = \int F({\bm x}, {\bm c}, I, t) I^{2/\delta}  dI,
\end{split}
\label{f1f2Evol}
\end{align}
where $f_1$ is related to the translational energy and  $f_2$ with the rotational energy dynamics  
 \citep{rykov1975model, andries2000gaussian}. The moments of reduced distribution $f_1({\bm x}, {\bm c}, t)$ are 
then same as the moments of single-particle distribution function 
\begin{align} 
\{ \rho, \rho {\bm u}, E_T \} = \left\langle \left\{ 1, {\bm c}, \frac{c^2}{2} \right\}, f_1 \right\rangle.
\end{align}
By construction, we have  the zeroth moment of $f_2({\bm x}, {\bm c}, t)$ as the 
rotational energy
\begin{align}
 E_{R}= \left\langle \left\langle  \frac{c^2}{2} + I^{2/\delta} , F\right \rangle \right \rangle- \left\langle f_1, \frac{c^2}{2} \right 
\rangle=  \left\langle \left\langle F, I^{2/\delta} \right\rangle\right\rangle\equiv \left\langle f_2, 1 \right \rangle = \frac{\delta}{2}  \rho \theta_R.
\label{f1f2Moments}
\end{align}
In other words, the temperature $\theta$ consists of 
contributions from the translational and rotational temperatures, and they follow 
the relation
\begin{equation}
 \theta = \left(\frac{3}{3+\delta} \right) \theta_T 
       +  \left(\frac{\delta}{3+\delta} \right) \theta_R,
 \label{thetadef}
\end{equation}
and in thermodynamic equilibrium the equipartition of energy requires    $\theta_R=\theta_T$. The heat flux for a  
polyatomic gas is $q_\alpha=  q^T_{\alpha} + q^R_{\alpha}$ where
 $q^T_{\alpha}$ is the translational heat flux and  $q^R_{\alpha}$ is
an additional heat flux due rotational energy.
The rotational heat flux  and a stress like quantity (second moment) are defined as
\begin{equation}
q^{\rm R}_{\alpha}=  \int d\xi f_2 \xi_\alpha, \quad
\sigma^{R}_{\alpha \beta}  = 
 \int d\xi f_2 \xi_\alpha \xi_\beta-\rho \theta^2\delta_{\alpha \beta}.
\end{equation}
Like the monoatomic gas, the evolution equation for this distribution function
$F({\bm x}, {\bm c}, t, I)$ with collisional kernel $\Omega(F,F)$  in the
Boltzmann form  is
\begin{align}
\begin{split}
 \partial_t F  + c_\alpha \partial_\alpha F = \Omega(F,F),
\end{split}
\label{FKineticEquation}
\end{align}
  which   is consistent with the equipartition of energy at equilibrium 
   \citep{pullin1978kinetic, kuvsvcer1989model}. Similar to the monoatomic gas, 
  one  defines the BGK collision kernel in terms of the two-particle distribution function
  for a polyatomic gas as  \citep{brull2009ellipsoidal}
\begin{align}
\begin{split}
\Omega_{\rm{BGK}} &= \frac{1}{\tau} \left(  F^{\rm MB}(\rho,\bm u, \theta , 
I) - F \right),\\
 F^{\rm MB}(\rho,\bm u, \theta , I)  &= 
 \frac{\rho \Lambda_{\delta} }{\left(2 \pi \theta \right )^{3/2} \theta^{\delta/2}} \exp{\left(-\left(\frac{({\bm c}-{\bm u})^2}{2 
\theta}  + \frac{I^{2/\delta}}{\theta} \right)\right)},
\end{split}
\label{BGKFCollision}
\end{align}
with normalisation factor $\Lambda_{\delta} = \int \exp{(-I^{2/\delta}) dI} $.
  Equation \eqref{FKineticEquation} with $\Omega_{\rm{BGK}}$ is written as two 
 kinetic equations for the reduced distributions $f_1({\bm x}, {\bm c}, t)$ and 
 $f_2({\bm x}, {\bm c}, t)$  by multiplying with $1$ and $I^{2/\delta}$ and then integrating over the internal energy variable as
\begin{align}
\begin{split}
 \partial_t f_1  + c_\alpha \partial_\alpha f_1 & = \frac{1}{\tau} \left(  
f_1^{\rm MB}(\rho,\bm u, \theta) - f_1 \right),\\
 \partial_t f_2  + c_\alpha \partial_\alpha f_2 & =  \frac{1}{\tau} \left( \frac{\delta}{2}\theta 
f_1^{\rm MB}(\rho,\bm u, \theta) - f_2 \right).
\end{split}
\label{f1f2Evolutuion}
\end{align}
This approach, where two reduced distributions are weakly coupled via temperature, recovers all the features of Eq.\eqref{BGKFCollision} and is widely adopted for polyatomic gases.  
\cite{andries2000gaussian} extended this approach via an extended  ES--BGK collision kernel    as
\begin{equation}
 \Omega_{\rm ESBGK}(F) = \frac{ Z_{\rm ES}}{\tau} 
\left( F^{\rm ES} (\rho, u, \lambda_{ij}, \theta_{\rm rel} ) - F \right),
\end{equation}
where $\lambda_{ij} = (1-\alpha) \left[ (1 -b) \theta_{T} \delta_{ij}+  b \Theta_{ij} \right]  + \alpha \theta \delta_{ij}$ 
 with a generalized Gaussian
$F^{\rm ES}$  
\begin{equation}
F^{\rm ES}(\rho, u, \lambda_{ij}, \theta_{\rm rel} ) = 
\frac{\rho \Lambda_{\delta} }{\theta_{\rm rel}^{\delta/2} \sqrt{\rm {det}[2 \pi \lambda_{ij}]}} 
\exp\left({-\frac{1}{2} \xi_i \lambda^{-1}_{ij} \xi_j} - \frac{I^{2/\delta}}{\theta_{\rm rel}} \right),
\end{equation}
 with $\theta_{\rm rel} = \alpha \theta + (1 - \alpha) 
\theta_{R}$ and $ Z_{\rm ES} = 1/(1 - b + b \alpha)$.
Similar to the monoatomic ES--BGK model, the parameter $b$ is used to tune the Prandtl number, while parameter $\alpha$ is used to tune the bulk viscosity coefficient independently. The reduced description which  generalizes  the ES--BGK model  in terms of the $f_1$ and $f_2$ is  
\begin{align}
\begin{split}
 \partial_t f_1  + c_\alpha \partial_\alpha f_1 & = \frac{ Z_{\rm ES}}{\tau} \left( f^{\rm ES}(\rho, u, \lambda_{ij}) - f_1 \right),\\
 \partial_t f_2  + c_\alpha \partial_\alpha f_2 & =  \frac{Z_{\rm ES}}{\tau} \left( \frac{\delta}{2} \theta_{\rm rel} f^{\rm ES}(\rho, u, \lambda_{ij}) - 
f_2 \right).
\end{split}
\label{andreiesTwoPopulation}
\end{align}
 A common feature between Eqs. \eqref{f1f2Evolutuion} and \eqref{andreiesTwoPopulation} is that the kinetic equation for translation distribution function $f_1$ is coupled with the kinetic equation for the rotational distribution function $f_2$ via rotational temperature $\theta_{R}$ only.  At this point, it might be instructive to analyse the moment chains
 of the BGK and ES--BGK systems. As both the
 collision kernels conserve mass and momentum, the evolution equations for  density and momentum are of the same form as that of monoatomic gas. In particular
\begin{align}
\begin{split}
 \partial_t \rho + \partial_{\alpha} (\rho u_{\alpha}) &=0, \\
 \partial_t (\rho u_{\alpha}) + \partial_{\beta} \left(\rho u_{\alpha}
 u_{\beta}+ \rho \theta 
\delta_{\alpha \beta}+\hat{\sigma}_{\alpha \beta}\right)  &=0,
\end{split}
\label{massAndMomentum}
\end{align}   
  where $\hat{\sigma}_{\alpha \gamma} = \sigma_{\alpha 
\gamma} + \rho \left( \theta_T - \theta \right) \delta_{\alpha \gamma}$ 
is the modified stress tensor and the velocity evolution is 
\begin{equation}
 \partial_t  u_{\alpha}+u_{\beta}\partial_{\beta}  u_{\alpha}
 +  \frac{1}{\rho}\partial_{\beta} \left(  \rho \theta 
\delta_{\alpha \beta}+\hat{\sigma}_{\alpha \beta}\right)  =0.
\label{velocityStd}
\end{equation}
For polyatomic gases, the energy  equation gets an additional contribution from the rotational energy. 
In both the BGK and ES--BGK models,  the evolution equation for translational part of the energy and the rotational part of the energy $E_{R}=\delta \rho \theta_{R}/2$ are of the form 
\begin{align}
\label{trans_dia}
\begin{split}
 \partial_t  E_T  +  
 \partial_{\alpha} \left[ \left( E_T +  \rho \theta \right) u_\alpha  
 +  q_\alpha^{T} + u_\gamma 
\hat{\sigma}_{\alpha \gamma}\right] &= \frac{Z_E}{\tau} \frac{3 \rho}{2}    \left( \theta -  \theta_T  \right) ,
\\
 \partial_t  E_R
 + \partial_\alpha \left( E_R u_\alpha + q^R_\alpha\right) 
& =  \frac{Z_E}{\tau} \frac{\delta \rho }{2}    \left( \theta - \theta_R \right),
  \end{split}
  \end{align}
 with  $Z_E = 1 $  for the BGK model and
 $Z_E = \alpha Z_{\rm ES}$ for the ES--BGK model.
The evolution equation for the total energy is 
written as the sum of Eqs. \eqref{trans_dia} as
\begin{equation}
 \partial_t \left( E_T + E_R \right) +
 \partial_{\alpha} \left[ \left( E_T + E_R +  \rho \theta \right) u_\alpha
 +  q_\alpha + u_\gamma \hat{\sigma}_{\alpha \gamma}\right] = 0,
 \label{eq:TotalEnergy}
\end{equation}
where  the relationship between translational, rotational temperatures
(Eq.\eqref{thetadef}) is used to show that energy is collisional invariant.
 From Eqs. \eqref{f1f2Evolutuion} and \eqref{andreiesTwoPopulation},
 the stress evolution and the translational heat flux evolution equations
 in explicit form are 
\begin{align}
\begin{split}
\partial_t \sigma_{\alpha \beta}
&+ \partial_\gamma \left(u_\gamma  \sigma_{\alpha \beta} \right)
+ \partial_\gamma \overline{Q_{\alpha \beta \gamma}} 
+ \frac{4}{5} \overline{\partial_\beta  q^{T}_\alpha}
+ 2 \rho \theta_T \overline{\partial_\beta u_\alpha}
+ 2 \overline{\partial_\gamma u_\alpha \sigma_{\gamma \beta}}
= -  \frac{1} {\tau} \sigma_{\alpha \beta},\\
& \partial_t q^T_\alpha 
+ \partial_\beta \left(u_\beta q^T_\alpha \right)
+ \overline{Q_{\alpha \beta \gamma}} \partial_\beta u_\gamma
+ \frac{1}{2} \partial_\beta R_{\alpha \beta} + \frac{7}{5}  q^T_\beta
\partial_\beta u_\alpha
    + \frac{2}{5} q^T_{\alpha} \partial_\eta u_\eta
    + \frac{2}{5} q^T_{\beta} \partial_\alpha u_\beta \\
&- \frac{5}{2} \theta_T \partial_\alpha (\rho \theta_T)
  - \frac{\sigma_{\alpha \beta}}{\rho} \partial_\beta(\rho \theta_T)
 - \frac{5}{2} \theta_T  \partial_\kappa \sigma_{\kappa \alpha}
 - \frac{\sigma_{\alpha \beta}}{\rho} \partial_\kappa \sigma_{\kappa \beta}
= -\frac{Z_{q}}{\tau} q^T_\alpha,
\end{split}
\end{align}
with $Z_q = 1$ for BGK and
 $Z_q = Z_{\rm ES}$ for the ES--BGK model.
Multiplying rotational energy equation from Eq.\eqref{trans_dia} with 
$u_\alpha$ and using Eq.\eqref{velocityStd} we have
\begin{align}
\begin{split}
\partial_t \left( E_R u_\alpha \right)
+  \partial_\beta \left( E_R u_\alpha u_\beta \right)
+ u_\alpha \partial_\beta q^{R}_\beta  
+ \frac{E_R}{\rho} \partial_\beta \left( \rho \theta \delta_{\alpha \beta}
+ \hat{\sigma}_{\alpha \beta} \right)
=  \frac{Z_E}{\tau} \frac{\delta}{2}  \rho u_\alpha    \left( \theta - \theta_R \right),
\label{ER_u_alpha}
\end{split}
\end{align} 
using which the rotational heat flux evolution obtained as first moment of
$f_2$ dynamics  is
\begin{equation}
 \partial_t q^R_\alpha
 + \partial_\beta \left(u_\beta q^R_\alpha \right)
 + q^R_\beta  \partial_\beta u_\alpha 
 + \partial_\beta  \sigma^{R}_{\alpha \beta}
 +  \partial_{\alpha} \left(\frac{\delta}{2} \rho \theta^2 \right) 
 - \frac{\delta}{2} \theta_R \partial_{\beta} \left(  \rho \theta \delta_{\alpha \beta} + \hat{\sigma}_{\alpha \beta} \right) = -  \frac{Z_{q}}{\tau} q^R_\alpha,
\end{equation}
where $ \sigma^{R}_{\alpha \beta} = \int f_2 \xi_\alpha \xi_\beta$.
The Eqs. \eqref{massAndMomentum} and \eqref{eq:TotalEnergy} form the compressible
Navier--Stokes--Fourier equations for a polyatomic gas.
A Chapman-Enskog analysis shows that the Eq.\eqref{massAndMomentum} can be written in
familiar compressible Navier--Stokes equation form as
\begin{equation}
 \partial_t (\rho u_\alpha) + \partial_\beta (\rho u_\alpha u_\beta) + 
\partial_\alpha p - \partial_\beta \left( 2 \eta \overline{\partial_\beta 
u_\alpha} + \eta_b  \partial_\kappa u_\kappa  \delta_{\alpha \beta}\right)
= 0
\end{equation}
with the shear and bulk viscosities
\begin{equation}
 \eta = p \tau, \quad \quad 
 \frac{\eta_b}{\eta} = \frac{2 \delta}{3(3+\delta)Z_E}.
\end{equation}

Similarly, an analysis of the translational and rotational
heat flux dynamics at $O({\rm Kn})$ leads to
$q^T_\alpha = -\kappa_T \partial_\alpha \theta , \,
q^R_\alpha = - \kappa_R \partial_\alpha \theta,$ with the translational
and rotational thermal conductivities $\kappa_T = 5p\tau/(2Z_q), \, \kappa_R = {\delta p\tau}/(2Z_q)$ respectively.
The effective thermal conductivity $\kappa = \kappa_T + \kappa_R 
= {(5 + \delta) p\tau/(2Z_q)} $
Thus, the Prandtl number is ${\rm Pr} = Z_q$ i.e., ${\rm Pr} = 1$ for BGK model and ${\rm Pr} = 1/(1-b+b\alpha)$ for the ES--BGK model.


\section{Reduced ES--BGK model for polyatomic gases}
\label{sec:ES--BGK-diatomic}
In the kinetic theory of gases, one often builds an extended moment system in terms of physically relevant lower-order moments  \citep{grad1958principles}. 
In this spirit of Grad's moment method, one may ask  
whether a reduced description for rotational degrees of freedom is feasible.
It should be noted that the evolution equation of $f_1$ is only weakly coupled
with the evolution of $f_2$  via  $\theta_R$.
An appropriate choice in the current context is a reduced description in terms
of lower-order moments of second distribution $f_2$. 
For example, the rotational component can be modelled by the evolution of
two scalars -- rotational energy and its flux (which are the zeroth and the
first moment of $f_2$).   Such a class of reduced-order kinetic models might
be   better suited for large-scale hydrodynamic simulations. An extended BGK model for diatomic gases was formulated by \cite{kolluru2020extended} wherein the BGK collision model  was coupled with the rotational part of energy (zeroth moment of  $f_2$) which in itself follows an advection-relaxation equation.

%
%
We extend this approach by a generalized ES-BGK model for
polyatomic gases with tunable Prandtl numbers where the collision term is a linear combination of ES--BGK and the BGK collision kernels. In this model, the ES--BGK term describes relaxation to a temperature $\theta_T$ over a time $\tau$ whereas the BGK collision kernel describes relaxation to a temperature $\theta$ over a  time $\tau_1$. The kinetic equation of the unified model along with the evolution equation for the rotational energy   is  
\begin{align}
\begin{split}
 \partial_t f_1  &+  c_\alpha \partial_\alpha f_1  = \frac{1}{\tau} 
 \left( f^{\rm ES}(\rho ,\bm u,  \theta_T \delta_{\alpha \beta} + b 
\sigma_{\alpha \beta} ) - f_1\right) + \frac{1}{\tau_1} \left( f^{\rm 
MB}(\rho, \bm u, \theta ) 
 - f_1  \right),\\
\partial_t\left( E_R \right) 
&+ \partial_\alpha \left( E_R u_\alpha + q^{R}_\alpha \right) = 
\frac{1}{\tau_1} \left( \frac{\delta}{2} \rho\theta  
- E_R \right),
\end{split}
\label{eq:kinEqBGK}
\end{align}
with the form of heat flux due to internal degrees of freedom as
\begin{equation}
q^{R}_\alpha = - \kappa_R \partial_\alpha \theta_R.
\label{eq:rotHeatFlux_simple}
\end{equation}

This model is a minimal extension of the monoatomic ES--BGK model needed for modeling polyatomic gases which also recovers all important features such as the positivity, macroscopic limit, and the $H$ theorem.  Here, the Prandtl number is a tunable parameter due to the presence of two relaxation time scales, whereas the rotational part of the internal energy alters the specific heat ratio to that of a polyatomic gas.
The model satisfies the $H$ theorem, thus ensuring convergence to a unique equilibrium state.   A few important characteristic of the present model are as follows: 
\begin{itemize}
\item{\bf Conservation Laws}: The mass and momentum conservation equations for the proposed model are obtained 
by taking the zeroth and first moments of $f_1$ evolution 
(Eq.\eqref{eq:kinEqBGK}). 
The second moment signifying translational energy evolution equation is
\begin{equation}
 \partial_t E_T +
 \partial_{\beta} \left[ \left( E_T +  \rho \theta_T 
\right)u_\beta  + \sigma_{\beta\gamma}u_\gamma +  q^{T}_{\beta} \right] = 
\frac{\rho}{\tau_1} \left[ \frac{3}{2}\theta -  \frac{3}{2}\theta_T \right],
\label{keES--BGKeq1}
 \end{equation}
 which when combined with the rotational energy equation shows that the total energy is conserved. This implies that the  evolution equation for slow moments    
\begin{equation}
 M_{\rm slow} = \left\{\rho, \rho u_\alpha, \frac{1}{2} \rho u^2 
 + \left(\frac{3+\delta}{2}\right) \rho \theta) \right\},
\end{equation}
mass density, momentum density, and total energy density are
 \begin{align}
\begin{split}
 \partial_t \rho + \partial_{\alpha} (\rho u_{\alpha}) &=0, \\
 \partial_t (\rho u_{\alpha}) + \partial_{\beta} \left(\rho u_{\alpha}
 u_{\beta}+ \rho \theta 
\delta_{\alpha \beta}+\hat{\sigma}_{\alpha \beta}\right)  &=0,\\
\partial_t \left( E_T+E_R \right) +   
\partial_{\beta} \left[ \left( E_T+E_R+\rho\theta \right) u_\beta + 
\hat{\sigma}_{\beta\gamma}u_\gamma +  q_{\beta} \right] &= 0.
\end{split}
\label{ConservedMomEq}
\end{align} 
Thus, the conservation laws have correct macroscopic form. 
\item{\bf H--Theorem}: 
For polyatomic gases, a part of entropy production should be due to rotational degrees of freedom. In the current model, as internal degrees of freedom are accounted for in a mean-field manner, similar to the Enskog equation one would expect that entropy contribution should only depend on rotational energy  \citep{resibois1978h}. Thus, we write generalized $H$--function for polyatomic gas $H_1$ in Sackur--Tetrode form as a sum of Boltzmann part for monoatomic contribution and rotational part $k \rho \ln \theta_R$  
 \citep{huang2009introduction}  
\begin{equation}
 H_1 = H + k \rho \ln \theta_R,
\end{equation} 
 with $k$ being an unknown scale factor to be fixed later.
 On multiplying Eq.\eqref{eq:kinEqBGK} with $\ln f$, and integrating over 
the velocity space we obtain the evolution of $H$ as
\begin{equation}
 \partial_t H + \partial_\alpha J_\alpha^H = \Sigma^{\rm ESBGK} + \frac{\tau}{\tau_1}\Sigma^{\rm BGK} \left( f^{\rm MB} (\rho, {\bm u}, \theta) \right) - \frac{3\rho}{2\tau_1} \frac{\theta-\theta_T}{\theta},
 \label{husual}
\end{equation}
where $J_\alpha^H$ is related to the entropy flux with $\Sigma^{\rm ESBGK}, \Sigma^{\rm BGK}$ being the entropy production due to the ES--BGK and the BGK terms respectively. 
The evolution of the rotational energy (second equation in Eq.\eqref{eq:kinEqBGK}) can be rewritten   as 
\begin{equation}
 \partial_t \ln \theta_R + u_\alpha \partial_\alpha \ln \theta_R 
 + \frac{2}{\delta \rho \theta_R} \partial_\alpha q^R_\alpha
 = \frac{ \theta - \theta_R }{\tau_1 \theta_R}.
\end{equation}
Multiplying the above equation with $\rho$ and exploiting continuity we obtain
\begin{equation}
 \partial_t \left( \rho \ln \theta_R \right) +
 \partial_\alpha \left( \rho u_\alpha \ln \theta_R 
 + \frac{2}{\delta} \frac{q^R_\alpha}{\theta_R} \right) 
 = 
 \frac{\rho}{\tau_1} \frac{ \theta - \theta_R}{\theta_R} 
 - \frac{2}{\delta} \frac{q^R_\alpha}{\theta_R^2}  \partial_\alpha \theta_R.
\label{hrotkin}
\end{equation}

Thereby, adding Eq.\eqref{husual} and Eq.\eqref{hrotkin} and using the form of $q^R_\alpha$ from Eq.\eqref{eq:rotHeatFlux_simple} the evolution of $H_1$ is  obtained as
\begin{equation}
\partial_t H_1 + \partial_\alpha \left(J_\alpha^H + k \rho u_\alpha \ln \theta_R 
+ k \frac{2}{\delta}\frac{q^R_\alpha}{\theta_R} \right) 
= \Sigma^{\rm ESBGK} + \Sigma^{\rm BGK} + \hat\Sigma,
\end{equation}
where the right hand side is the net entropy production with contributions 
from the ES--BGK collision, the BGK collision, and the rotational component of 
the model. Here,
\begin{equation}
\hat\Sigma = -\frac{3\rho}{2\tau_1} \frac{\theta-\theta_T}{\theta}  
+ \frac{k \rho}{\tau_1 } \frac{\theta - \theta_R }{\theta_R} 
+ k \frac{2}{\delta}\kappa_R \left( \frac{\partial_\alpha \theta_R}{\theta_R} \right)^2.
\label{omega1}
\end{equation}
Similar to the standard BGK or ES--BGK case \citep{andries2000gaussian},
the entropy production  $\hat\Sigma$ in this model is non-positive too. 
This is achieved by choosing $k=-\delta/2$ and exploiting the relation
Eq.\eqref{thetadef} 
to rewrite $\hat\Sigma$ as 
\begin{equation}
\hat\Sigma = - \frac{\delta}{2} \frac{\rho}{\tau_1} \frac{(\theta-\theta_R)^2}{\theta \theta_R}
-  \kappa_R \left( \frac{\partial_\alpha \theta_R}{\theta_R} \right)^2 \leq0.
\label{finalOmega1}
\end{equation}
Hence, the proposed model satisfies the $H$ theorem.

\item{\bf Hydrodynamics}:
In order to derive the hydrodynamic limit and the transport coefficients, the moments are typically categorized
into fast $M_{\rm fast}$ and slow moments  $M_{\rm slow}$.
The stress tensor and the heat flux  constitutes the relevant
set of fast moments along with the translational and rotational temperatures 
as they are not conserved
\begin{equation}
 M_{\rm fast} = \left\{\theta_T, \theta_R, \sigma_{\alpha \beta}, q_{\alpha} 
\right\}.
\end{equation}
The base state  is obtained from zero of collision from Eq.\eqref{eq:kinEqBGK} as 
\begin{equation}
f=f^{\rm MB}\implies \theta = \theta_T \quad {\rm and} \quad \theta = \theta_R.
\end{equation}
Thus, the fast moment can be expanded  around their equilibrium values in a   series   as
\begin{equation}
 M_{\rm fast} =  M_{\rm fast}\left(f^{\rm{MB}}\right)  +
 \tau M_{\rm fast}^{(1)} + \cdots.
\end{equation}
In Chapman--Enskog expansion, the time derivative of any quantity $\phi$ is expanded as 
\begin{align}
\begin{split}
\partial_t \phi &= \partial_t^{(0)} \phi + \tau \partial_t^{(1)} \phi + {\cal O}
(\tau^2).
\label{timeder}
\end{split}
\end{align}

The set of conservation laws (Eqs.\eqref{ConservedMomEq}) upon substituting the expansion of time derivative provide the definition of time derivative at ${\cal O}(1)$ of slow variables as Euler equations
 \begin{align}
\begin{split}
 \partial_t^{(0)} \rho + \partial_\alpha \left( \rho u_\alpha \right) &= 0,\\
 \partial_t^{(0)} \left( \rho u_\alpha \right) + \partial_\beta \left( \rho
 u_\alpha u_\beta + \rho \theta \delta_{\alpha \beta}  \right) &= 0,\\
\partial_t^{(0)} E + 
\partial_\beta \left( E u_\beta + \rho \theta u_\beta \right) &= 0.
\end{split}
\label{zerothOrder}
\end{align}
Thus, pressure  evolution at $O(1)$  satisfies the adiabatic  condition 
 for a polyatomic gas
\begin{equation}
 \left( \partial_t^{(0)} +  u_{\beta} \partial_{\beta} \right) 
 \left( \frac{p} {\rho^\gamma} \right) = 0, \text{ where } \gamma = \frac{5+\delta}{3+\delta}.
\end{equation}
Similarly, at order ${\cal O}(\tau)$ we have
\begin{align}
\begin{split}
 \partial_t^{(1)} \rho &= 0, \\
 \partial_t^{(1)} \left( \rho u_\alpha \right) + \partial_\alpha (\rho 
\theta^{(1)}_T) + \partial_\beta \sigma_{\alpha \beta}^{(1)} &= 0,\\
\partial_t^{(1)} E +  \partial_\beta  \left( \sigma_{\alpha 
\beta}^{(1)} u_\alpha + \rho \theta^{(1)}_T u_\beta + q_\beta^{(1)}\right) &= 
0.
\end{split}
\label{firstOrder}
\end{align}

The expressions for $\rho \theta_T^{(1)}$ and $\sigma^{(1)}_{\alpha \beta}$ 
can be obtained from the evolution equations of translational temperature 
(Eq.\eqref{translationalTemp}) and stress tensor (Eq.\eqref{stressEvolution})
respectively as [Details of derivations in Appendix \ref{evolEqns}] 
\begin{align}
\begin{split}
 \rho \theta_T^{(1)} = -\frac{2 \delta}{3(3+\delta)} \frac{\tau_1}{\tau} p  
\partial_\gamma u_\gamma, \quad
 \sigma^{(1)}_{\alpha \beta} = -\frac{2 p \overline{\partial_\beta 
u_\alpha}} {B-b},
\end{split}
\end{align}
where $B = 1 + \tau/\tau_1$. Substituting the above expressions in momentum conservation equation,  $O(\tau)$  hydrodynamics with shear viscosity $\eta$ and bulk viscosity $\eta_b$ is
\begin{align}
\begin{split}
 \eta = \frac{p \tau}{B-b}, {\quad \rm and} \quad
 \eta_b =  \frac{2 \delta}{3(3+\delta)}  p \tau_1.
\end{split}
\label{eq:viscosities}
\end{align}

Similarly, translational thermal conductivity for the model is obtained
from the translational heat flux evolution (Eq.\eqref{translationalHeatFluxEvolutionEquation}).
A Chapman--Enskog expansion of Eq.\eqref{translationalHeatFluxEvolutionEquation} by substituting 
$ q^T_\alpha = (q^T_\alpha)^{\rm{MB}}  + \tau  {(q^T_\alpha)}^{(1)} + {\cal O}(\tau^2)$
at ${\cal O}(1)$ yields 
\begin{align}
\begin{split}
 \left(1 +  \frac{\tau}{\tau_1} \right) {(q^T_\alpha)}^{(1)} = 
-\frac{5}{2} \rho \theta \partial_\alpha \theta.
\end{split}
\end{align}
Thus, the translational thermal conductivity is 
$\kappa_T = 5 p \tau/(2B)$ which means  the total thermal conductivity 
$\kappa = \kappa_T + \kappa_R$ with $k_r = \kappa_R/\kappa_T$ is $\kappa =  {5 p \tau }/(2B)  \left(1  + k_r \right)$. 
Hence, 
\begin{equation}
{\rm Pr} = \frac{\eta C_p}{\kappa} =  \left(\frac{B}{B-b} \right)
\left(1+\frac{\delta}{5}\right)  \left(\frac{1}{1+k_r} \right).
\end{equation}

The relaxation times $\tau$ and $\tau_1$ are fixed 
based on the shear and bulk viscosities respectively and the free parameters 
$b$ and $k_r$ can be adjusted to obtain a target Prandtl number.
\end{itemize}


\section{Discretizing via lattice Boltzmann Method}\label{sec:numMtd}

In this section, we formulate a lattice Boltzmann  scheme for solving the proposed model. 
Firstly, the velocity space is discretized into a discrete velocity set  ${\bm c} = \{ {\bm c}_i, i=1 \cdots N \}$ consisting of $N$ vectors. These vectors
form the links of a lattice that should satisfy appropriate isotropy conditions
\citep{succi2001lattice, atif2018higher}. In particular, we validate the proposed kinetic model using a 41-velocity crystallographic lattice from \cite{kolluru2020lattice}, which uses a body-centered cubic (bcc) arrangement of grid points. The bcc lattice contains two simple cubic lattices offset by half the length of grid spacing in all directions for better spatial discretization  \citep{namburi2016crystallographic}. The discrete velocities and their corresponding weights for this RD3Q41 model are given in Table \ref{tab:weights}.  The kinetic equation (Eq.\eqref{eq:kinEqBGK}) in discrete in velocity space 
for populations $f_{1i}$  is
\begin{align}
\begin{split}
 \partial_t f_{1i}\left({\bm x},{\bm c},t \right)  +  c_\alpha \partial_\alpha f_{1i}\left({\bm x},{\bm c},t \right)  = \Omega_{i} ({\bm x},t),
 \label{discrete_f1}
 \end{split}
\end{align}
where
\begin{align}
\begin{split}
\Omega_{i}\left({\bm x},t \right)  &= \frac{1}{\tau} 
 \left[ f^{\rm ES}_{1i}(\rho ,\bm u,  \theta_T \delta_{\alpha \beta} 
 + b \sigma_{\alpha \beta} ) - f_{1i} \right] + \frac{1}{\tau_1} \left[ f_{1i}^{\rm 
Eq}(\rho, \bm u, \theta) - f_{1i}  \right],
 \end{split}
\end{align}
with moments of the discrete populations $f_{1i}$ defined as
\begin{align}
\begin{split}
 \rho(f_{1}) = \sum_i f_{1i}, \;\; 
 \rho u_\alpha(f_{1}) =   \sum_i f_{1i} c_{i \alpha}, \quad 
 \theta_T (f_{1}) = \frac{1}{3\rho(f_{1})}
               \left( \sum_i f_{1i} c^2_{i} - \rho {\mathbf u}^2 \right).
 \end{split}
\end{align}

\begin{table} 
\begin{center}
\begin{tabular}{cccc} 
\hline \hline
 Discrete Velocities($c_i$) &  Weight($w_i$) \\ \hline
$\left( 0,0,0\right)$ &  $ \left(52 - 323 \theta_0 + 921 \theta_0^2 - 1036 \theta_0^3\right)/52$ \\
$\left(\pm 1, 0, 0  \right),\left(0, \pm 1, 0  \right),\left( 0, 0, \pm 1  
\right) $ &  $ \theta_0 \left(12 - 38 \theta_0 + 63 \theta_0^2\right)/39 $ \\
$\left(\pm 2, 0, 0  \right),\left(0, \pm 2, 0  \right),\left( 0, 0, \pm 2 
\right) $ &  $ \theta_0 \left(3 - 29 \theta_0 + 84 \theta_0^2\right)/312  $ \\
$\left(\pm 1, \pm 1, 0  \right),\left(\pm 1, 0, \pm 1  \right),  \left( 0,\pm 1,
\pm 1  \right)$ &  $  \theta_0 \left( 45 \theta_0 - 6 - 77 \theta_0^2\right)/26           $ \\
$\left(\pm 1, \pm 1,  \pm 1  \right)$ &   $ \theta_0 \left(20 - 163 \theta_0 + 378 \theta_0^2 \right)/312       $ \\
$\left(\pm 0.5, \pm 0.5,  \pm 0.5  \right)$ &  $ 8\theta_0 \left(4 - 17 \theta_0 + 21 \theta_0^2 \right)/39 $ \\ \hline 
\hline
\end{tabular}
\caption{Velocities and their corresponding weights for the RD3Q41 model
with $\theta_0 = 0.2948964908710633$}
\label{tab:weights}
\end{center}
\end{table}

To have a numerically efficient scheme for the $N$ coupled partial differential equations of Eq.\eqref{discrete_f1},  it is desirable to have large time steps, i.e., $\Delta t \gg \tau$. Upon integrating Eq.\eqref{discrete_f1} along the characteristics and approximating the integral related to collision term via trapezoid rule, we obtain the implicit relation
\begin{align}
f_{1i}({\bm x} + {\bm c} \Delta t, t + \Delta t) = f_{1i}( {\bm x}, t) + \frac{\Delta t}{2} \left[ \Omega_i ({\bm x},t) + \Omega_i ({\bm x} + {\bm c}t,t+\Delta t) \right],
\end{align}
which is made explicit  by a transformation to an auxiliary population $g_{1i}\left({\bm x},{\bm c},t \right)  = f_{1i} \left({\bm x},{\bm c},t \right) - ({\Delta t}/{2}) \Omega_{i} \left({\bm x},{\bm c},t \right)$. This implies the evolution equation for $g_{1i}$ is
\begin{align}
\begin{split}
 &g_{1i}\left({\bm x}+ {\bf c}_i \Delta t, t+\Delta t \right)  =
 g_{1i}\left({\bm x},t \right)  \left( 1 - 2\beta^* \right) +  2 \tau^* \beta^* \Omega_i\left({\bm x},{\bm c},t \right), 
\end{split}
\end{align}
with ${1}/{\tau^*} =  {1}/{\tau}  + {1}/{\tau_1}$ and $\beta^* = {\Delta t}/{(2 \tau^* + \Delta t)}$.
The moments of the auxiliary distribution $g_{1}$ are related to the moments of
discrete populations $f_{1}$ as
\begin{align}
\begin{split}
&\rho(g_{1}) = \rho(f_1), \; 
\mathbf{u}(g_1) = \mathbf{u}(f_1), \;
\theta_T(g_1) =  \theta_T(f_1) + 
  \left( \frac{\delta \Delta t}{2\tau_1 (3 +\delta )} \right) 
     \left( \theta_T(f_1) -  \theta_R \right) , \quad \\
& \sigma_{\alpha \beta}(g_1) = \sigma_{\alpha \beta}(f_1)  \left(1 + \frac{\Delta t}{2 \tau^*} - \frac{\Delta t}{2\tau} b  \right).
\end{split}
\end{align}

To solve the the second part of Eq.\eqref{eq:kinEqBGK} that represents the internal energy, we write 
\begin{equation}
 \frac{\partial \theta_{\rm Rot}}{\partial t } + u_\alpha \frac{\partial 
 \theta_{\rm Rot} }{\partial x_\alpha} = \frac{1}{\tau_1} \left(\theta - 
\theta_{\rm Rot} \right) 
+ \frac{2}{\rho \delta} \kappa_R \nabla^2 \theta_{\rm Rot},
\end{equation}
by exploiting the Eqs. \eqref{thetadef}, \eqref{eq:rotHeatFlux_simple}, and the continuity equation.
The above equation is an advection-relaxation-diffusion equation that is solved by the steps detailed below:
\begin{enumerate}
\item $\,$  The relaxation equation 
\begin{equation}
 \label{Erotrelaxation}
 \frac{\partial \theta_{\rm Rot}}{\partial t } = \frac{1}{\tau_1} 
\left(\theta - \theta_{\rm Rot} \right), 
\end{equation}
is first solved using the backward Euler method for a half-time step along with
the relation in Eq. \eqref{thetadef} to obtain 
\begin{equation}
  \theta^{t+\Delta t/2}_{\rm Rot} = \frac{1}{1+X} \theta^t_{\rm Rot}
  + \frac{X}{1+X}  \theta_T(f_{1}),
\end{equation}
  where $X = 3 \Delta t / (2 \tau_1 (3+\delta))$.
\item  $\,$ The MacCormack scheme  \citep{maccormack2003effect}, which uses forward and backward differences for spatial derivatives in the predictor and corrector steps respectively,  is used to solve the advection equation 
\begin{equation}
 \label{Erotadvection}
 \frac{\partial \theta_{\rm Rot}}{\partial t } + u_\alpha \frac{\partial 
 \theta_{\rm Rot} }{\partial x_\alpha} = 0.
\end{equation} 
\item  $\,$ The diffusion equation 
\begin{equation}
 \frac{\partial \theta_{\rm Rot}}{\partial t} = \frac{2}{\rho \delta} \kappa_R \nabla^2 \theta_{\rm Rot},
 \label{ErotDiffusion}
\end{equation}
is then solved using the standard forward time centered space (FTCS) scheme to get $\theta^{\rm dif}_{\rm Rot}$.
\item  $\,$ Finally, the second part of relaxation is completed by an advance of $\theta^{\rm dif}_{\rm Rot}$ by another half time-step $\Delta t/2$ leading to the final solution at
$t+\Delta t$ as
\begin{equation}
  \theta^{t+\Delta t}_{\rm Rot}  =  \frac{1}{1+X} \theta^{\rm dif}_{\rm Rot}
  + \frac{X}{1+X}  \theta_T(f_{1}).
\end{equation}
\end{enumerate}

Note that the  choice of the solver for the evolution equation
of rotational energy is independent of the lattice Boltzmann solver used for solving $f_{1}$.
In the next section, we validate the proposed numerical model   by simulating a few benchmark problems related to acoustics, hydrodynamics, and heat transfer such as propagation of an acoustic pulse, startup of a simple shear flow, thermal conduction, and viscous heat dissipation.

\section{Validation}\label{sec:validation}

\begin{figure}
 \centering
  \includegraphics[scale=0.5]{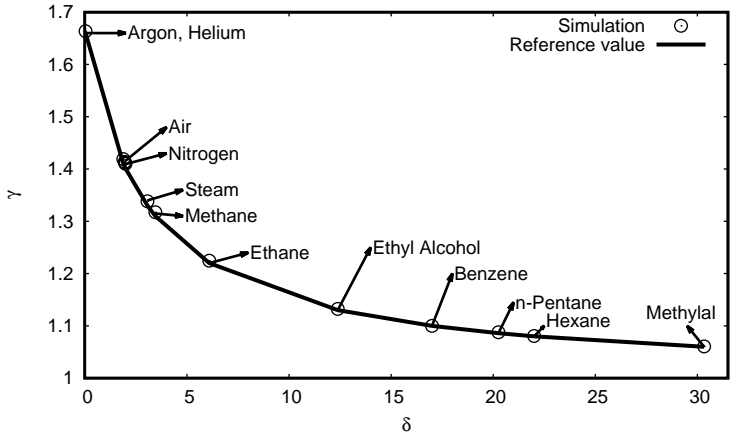} 
  \caption{Specific heat ratio in  simulating
  sound propagation in different gases. The line represents the reference value
  with $\delta$  the number of effective rotational degrees of freedom
  for various gases as listed in Table \ref{tab:effectiveDelta}}.
 \label{gamma_dof}
\end{figure}

\begin{figure}
 \centering
  \includegraphics[scale=0.5]{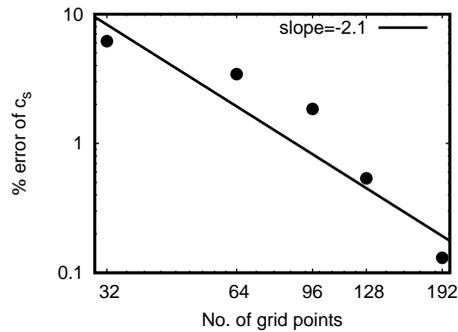} 
  \caption{Percentage error of speed of sound ($ | 1 - c_s^{\text{simulation}}/c_s^{\text{expected}}|\times100$) versus grid sizes
  showing a second order convergence for air.}
 \label{convergence}
\end{figure}

In this section, we validate the model by contrasting simulation result with various benchmark results. As a first example, we consider a periodic domain $\left[-\pi,\pi\right]$ with $128 \times 4 \times 4 $ lattice points to verify numerical  sound speed. We initialize the domain with  a pressure fluctuation of the form $p(x, t = 0) = p_0 (1.0 + \epsilon \cos(x))$ with $p_0 = \theta_0$.
The pressure pulse is expected to reach the same state as the
initial condition after one acoustic time period $(t_a)$.
The $L_2$-norm of the pressure fluctuation is computed using the current
state and initial state which is expected to be minimum when the waves
are in-phase. The number of time-steps taken to achieve the least $L_2$-norm is
used to compute $t_a$ and the speed of sound as $c_s = L/t_a$ where $L$ is
the domain length. The $\gamma= c^2_s/\theta_0$ is thus computed from the speed of sound and the lattice temperature.
Here, we demonstrate the versatility of the model by simulating several real fluids by imposing the effective rotational degrees of freedom $\delta$ as given in Table \ref{tab:effectiveDelta}. We show in Figure \ref{gamma_dof} that our model accurately recovers the specific heat ratio for various polyatomic
gases, even for fractional (effective) rotational degrees of
freedom. The proposed model remains accurate even for fractional rotational degrees of freedom, thereby achieving any target specific heat ratio values.
In Figure \ref{convergence}, we perform a grid convergence study for air
and observe a second order convergence.
We perform additional validation studies  by restricting our attention to
diatomic gases with variable Prandtl number.


\begin{table}
  \begin{center}
  \begin{tabular}{ccc}
  Fluid          &  $\gamma$  & Effective-$\delta$ \\  \\[3pt]
  Argon, Helium  &   1.66     & 0.03               \\
  Air            &   1.403    & 1.96               \\
  Nitrogen       &   1.404    & 1.95               \\
  Steam          &   1.33     & 3.06               \\
  Methane        &   1.31     & 3.45               \\
  Ethane         &   1.22     & 6.09               \\
  Ethyl alcohol  &   1.13     & 12.38              \\
  Benzene        &   1.1      & 17                 \\
  n-Pentane      &   1.086    & 20.26              \\
  Hexane         &   1.08     & 22                 \\
  Methylal       &   1.06     & 30.33              \\
\end{tabular}
\end{center}
\caption{Specific heat ratios of real fluids \citep{green2019perry} and their effective rotational
degrees of freedom}
\label{tab:effectiveDelta}
\end{table}

 Next,  we study the absorption of sound in a dissipative compressible medium. 
The presence of both viscosity and thermal conductivity leads to the dissipation of energy in the sound waves. 
For an emitted wave, the pressure perturbations $p'$ far away from the source decays during a finite time is
 \citep{landau1987fluid}
\begin{equation}
  p'(r,t) \propto \left({\rm La} \ rL \right)^{-1/2}
    \exp{\left(- \frac{\left(r - c_s t \right)^2}{2 {\rm La} \ rL  } \right)},
\end{equation}
 where the dimensionless Landau number ${\rm La}$ is
 \citep{ansumali2005thermodynamic}
\begin{equation}
 \rm{La} = \rm{Kn} \left(\frac{4}{3} + \lambda \right) 
 + \frac{\rm{Kn}}{\rm{Pr}} \left(\gamma - 1 \right).
\end{equation}
Here, $\lambda$ is the ratio of bulk to shear viscosities and $\rm{Kn}$ is the 
Knudsen number.
The form of pressure perturbation shows that the wave profile is Gaussian-like at
large distances and the width of the wave is proportional to $\sqrt{{\rm La}}$
for a fixed domain length $L$.

\begin{table}
  \begin{center}
\def~{\hphantom{0}}
  \begin{tabular}{cc}
  $\rm{Pr}$ & $\rm{La}$  \\[3pt]
  1.4  & 2.3302$\times 10^{-3}$  \\ 
  2.0  & 2.0311$\times 10^{-3}$ \\
  5.0  & 1.6124$\times 10^{-3}$ \\
  10.0 & 1.4729$\times 10^{-3}$ \\ 
  \end{tabular}
  \caption{Variation of $\rm{La}$ with $\rm{Pr}$ at ${\rm Kn} = 10^{-3}$.}
  \label{tab:kd}
  \end{center}
\end{table}

\begin{figure}
 \centering
  \includegraphics[scale=0.5]{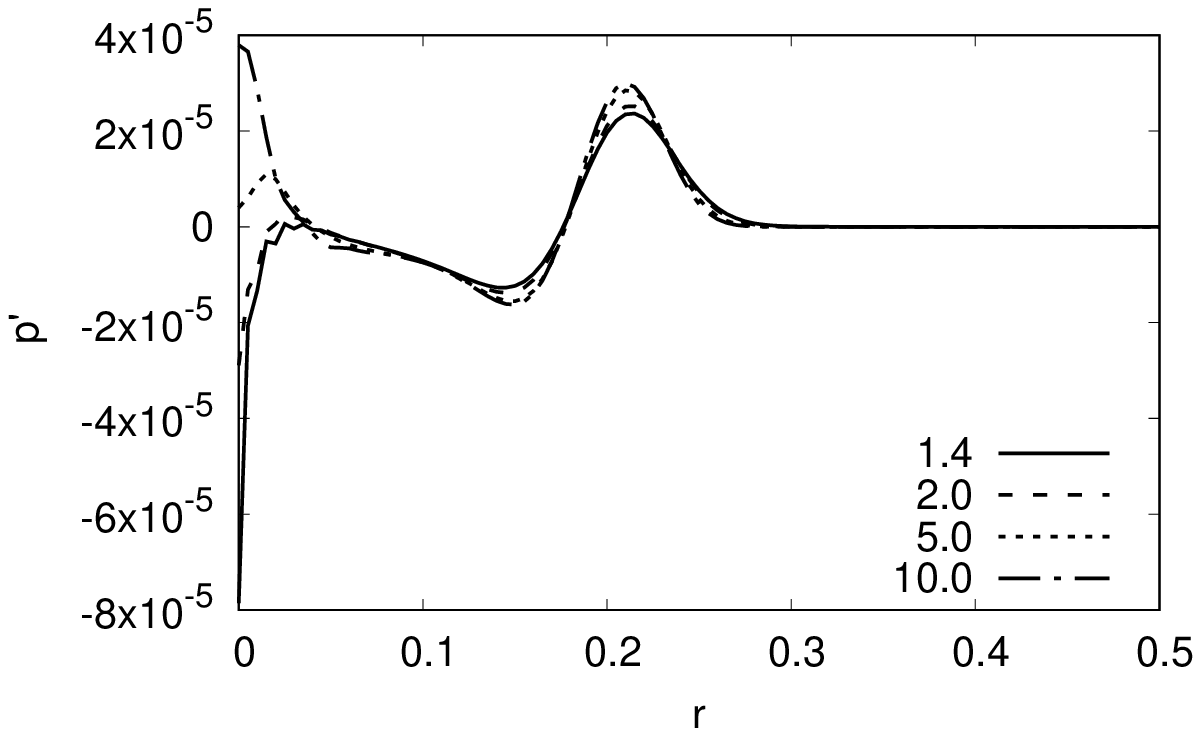} 
  \includegraphics[scale=0.5]{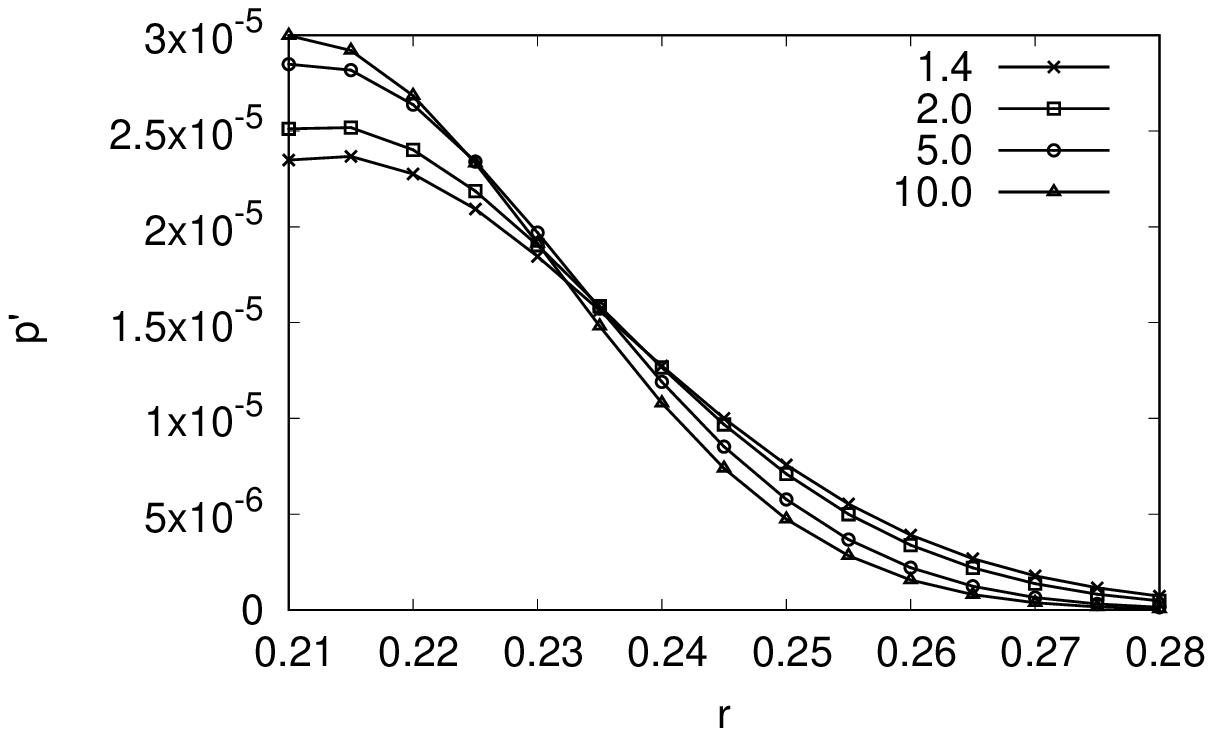} 
	\caption{Pressure perturbations versus Pr (zoomed in the right plot): the width of the Gaussian 
 wave increases with a decrease in $\rm{Pr}$.}
 \label{landau}
\end{figure}

To demonstrate  the  effectiveness of the LB scheme, we perform a simulation at a fixed $\rm{Kn}$ value of $10^{-3}$ on a domain of size $400 \times 400$
at Prandtl numbers $1.4$, $2, 5$, and $10$. We initialize the domain with a normal density perturbation of amplitude $0.001$ at the center of the fluid of uniform density $1.0$ at rest.
From Eq.\eqref{eq:viscosities} and setting $\tau = (3/5) \tau_1$ one obtains $\lambda = {224}/(225 \,{\rm Pr})$.
Using this above relation and $\gamma = 7/5$ the Landau  number $\rm{La}$ is calculated as 
\begin{equation}
 \rm{La} = \rm{Kn} \left(\frac{4}{3} +\frac{314}{225} \frac{1}{Pr} \right). 
\end{equation}
The Landau numbers for the chosen set of parameters are listed in
Table \ref{tab:kd}. It is evident that $\rm{La}$ is inversely proportional to $\rm{Pr}$  which suggests that the width of the Gaussian increases with a decrease in $\rm{Pr}$ number.
The pressure fluctuations far from the source of perturbation after $t=0.2 t_a$ for various Prandtl numbers are plotted in Figure \ref{landau}, where $t_a = L/c_s$ is the acoustic time scale. It is evident that as expected the width of the wave is inversely proportional to the Prandtl number.

Next, we  investigate the propagation of an acoustic pulse in  a diatomic gas with
$\gamma=7/5$ where the isentropic
speed of sound  is $c_s = \sqrt{\gamma \theta}$.
An axisymmetric pressure pulse is initialized at the centre of a domain
of size $[-1,1]$ with $256\times 256\times 4$ grid points. The acoustic pulse
is of the form
\begin{equation}  
p(x,y,t=0) = p_0 \left(1.0 + \epsilon e^{-\alpha r^2} \right),
\end{equation} 
with $p_0 = \theta_0$, $\epsilon = 0.001$,  $b = 0.1$, $\alpha = \ln(2)/b^2$, and $r=\sqrt{x^2 + y^2}$.
For low amplitudes of pressure fluctuations and low viscosity, 
the exact form of the pressure fluctuation is known as the solution of
the linearized Euler equations as ~ \citep{tam1993dispersion}
 \begin{equation}
p'(x,y,t) = p_0 \times \frac{\epsilon}{2 \alpha} \int_0^{\infty} 
\exp\left(\frac{-\xi^2}{4 \alpha}\right) \cos(c_s \xi t) J_0(\xi r) \xi\ d\xi,
\end{equation}
where $J_0$ is the Bessel function of the first kind of zero-order
 \citep{abramowitz1965handbook}.
\begin{figure}
 \centering
 \includegraphics[scale=0.5]{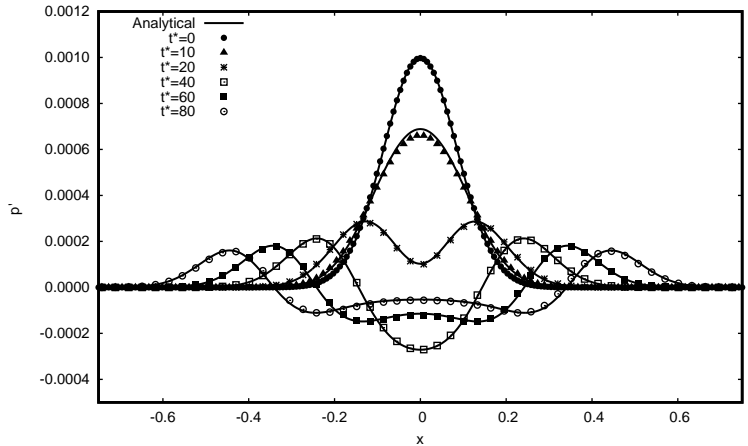} 
 \caption{Comparison of the pressure fluctuation along the 
 centerline at time $t^*$ from LB simulation(points) and exact solution(line).}
 \label{fig:2}
\end{figure}
 Figure. \ref{fig:2} shows that the pressure fluctuations from the simulation
and the exact solution  along the centerline of $y$-axis are in agreement.

\begin{figure}
\centering
 \includegraphics[scale=0.45]{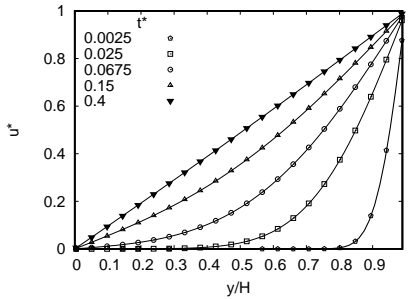}
\caption{ Transients in a planar Couette flow}
\label{fig:couette}
\end{figure}

Next, we simulate the transient hydrodynamics  in the startup of a simple shear flow between two flat plates separated by a distance $L$ on a grid of size $128 \times 64 \times8$ with diffusive wall boundary condition \citep{ansumali2002kinetic} and periodicity in the other two directions.
Here, the top plate is suddenly started with a velocity $u_w$ while the bottom
plate remains stationary. The  viscous effects play an important role in the
development of the flow which is driven by momentum diffusion.   Figure \ref{fig:couette} contrasts the solutions at 
various diffusion times $t^* = t/(L^2/\nu)$ obtained from our simulations
with the known analytical solution for the velocity 
 \citep{pozrikidis1997introduction}
\begin{equation}
 u^* = \frac{u}{u_w} =  \frac{y}{L} - \frac{2}{\pi} \sum_{k=1}^\infty \left[ 
 \frac{1}{k} \exp{\left( - k^2 \pi^2 \frac{\nu t}{L^2}\right)} \sin{\left(k \pi \left(1 - \frac{y}{L} \right) \right)} 
 \right].
\end{equation}
As expected, the simulation recovers the analytical solution with
good accuracy.

\begin{figure}
\centering
\begin{subfigure}{.5\textwidth}
  \centering
  \includegraphics[scale=0.45]{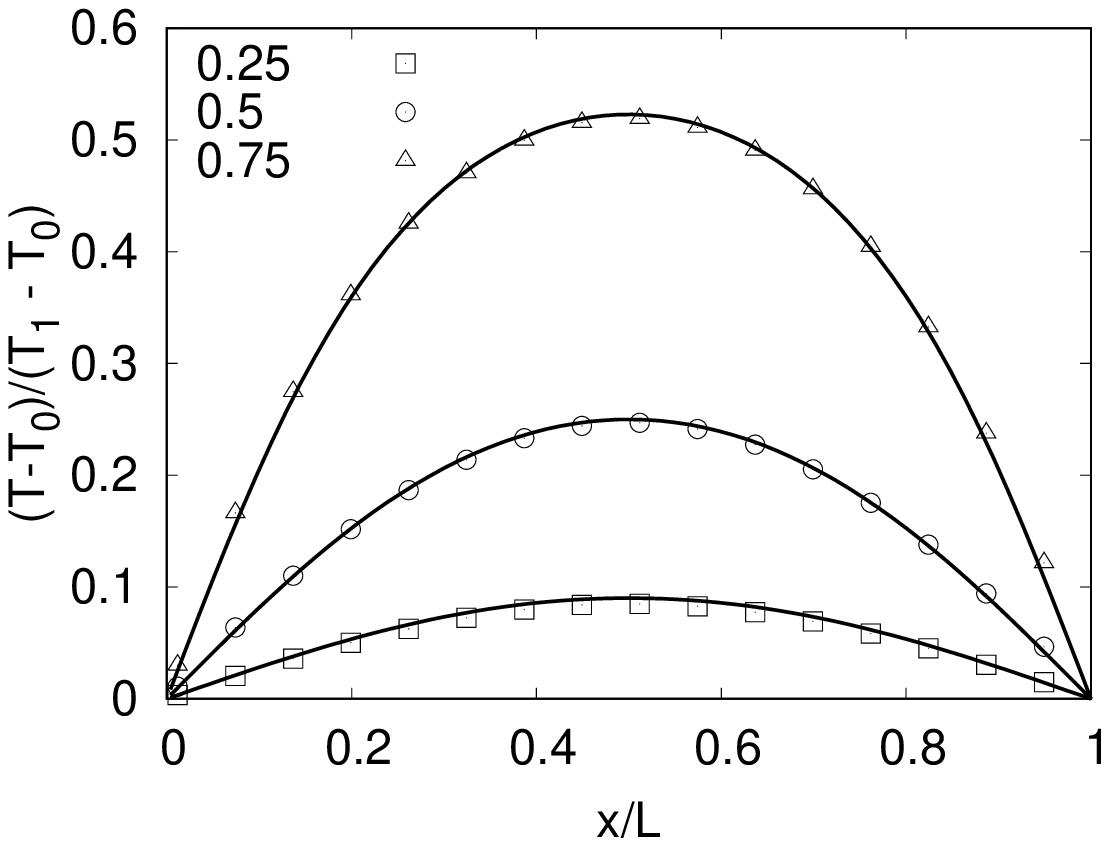}
\end{subfigure}%
\begin{subfigure}{.5\textwidth}
  \centering
  \includegraphics[scale=0.45]{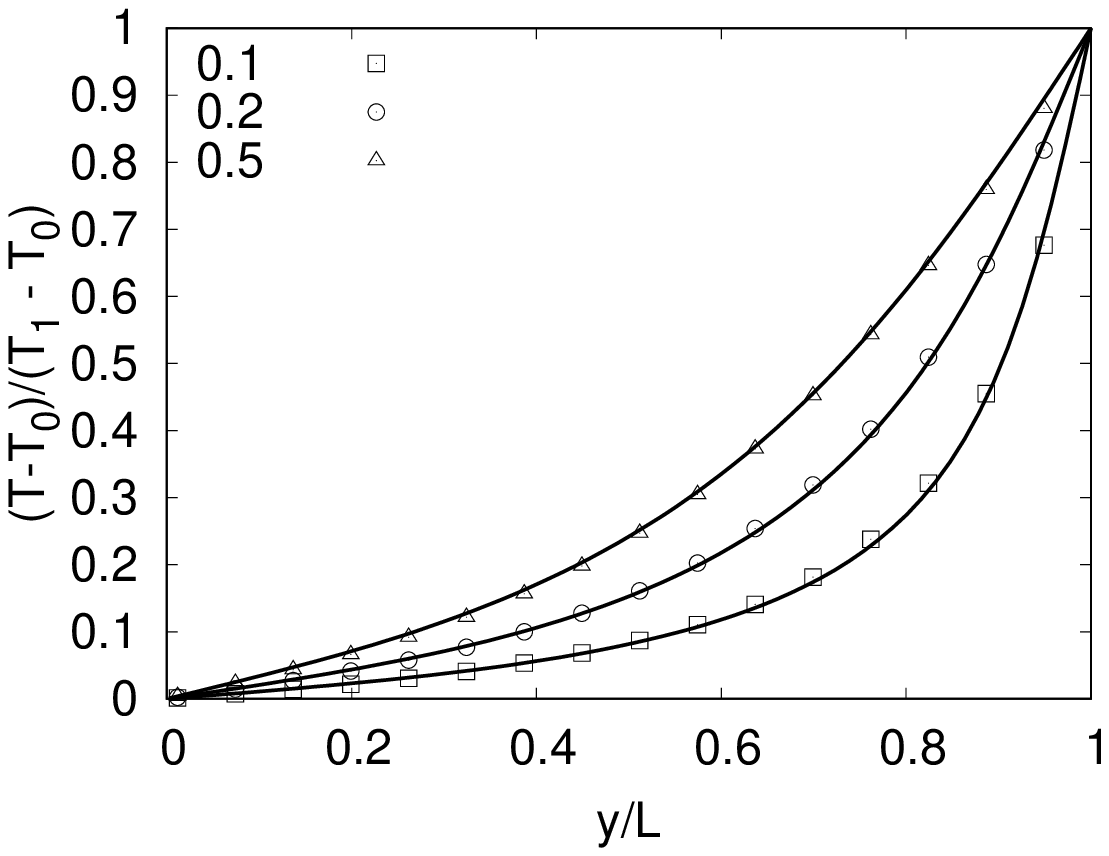}
\end{subfigure}
\caption{Temperature profiles at $y=0.25L,\ 0.5L$, and $0.75L$ (left)
and at $x=0.1L,\ 0.2L$, and $0.5L$ (right) in a 2D heated cavity.}
\label{fig:2Dconduction}
\end{figure}

Next, we investigate the effects of thermal conduction by considering a 
setup consisting of fluid confined in a square cavity of size $[L,L]$ with
$128\times 128$ points and stationary walls. The top wall maintained at a higher temperature $T_1$, while the other three walls are maintained at a temperature $T_0 (< T_1)$.  Diffusive wall boundary conditions are applied
in both directions. The analytical solution for the temperature profile at steady state is  \citep{leal2007advanced}
\begin{equation}
 \frac{T - T_0}{T_1 - T_0} = \frac{2}{\pi} \sum_{n=1}^{\infty} \frac{(-1)^{n+1}+1}{n} \sin\left({n\pi x}\right) 
\frac{\sinh(n\pi y)}{\sinh(n \pi)}.
\end{equation}
  Figure \ref{fig:2Dconduction} shows that the simulated temperature profiles along lines $x=0.1L,\ 0.2L$, and $0.5L$ and
along $y=0.25L,\ 0.5L$, and $0.75L$ for a temperature difference of $0.1 \theta_0$   matches well with the analytical solution.

Next, we validate our model for a thermal Couette flow problem
to evaluate its capability in simulating viscous heat dissipation at various
Prandtl numbers.
We study the steady-state of a flow induced by a wall at $y=H$ moving with a constant horizontal velocity $U_0$ and maintained at a constant 
elevated temperature $T_1$. The lower wall at $y=0$ is kept stationary
at a constant temperature $T_0$ ($T_1 > T_0$). The analytical solution for the temperature profile for this
setup is~ \citep{bird2015introductory}
\begin{equation}
 \frac{T - T_0}{\Delta T} = 
\frac{y}{H} + \frac{\rm Pr \ \rm Ec}{2} \frac{y}{H} \left(1 - \frac{y}{H} \right),
 \label{eckert}
\end{equation}
where $\Delta T = T_1 - T_0$ is the temperature difference between the 
two walls and ${\rm Ec} = U_0^2/(c_p \Delta T)$ is the Eckert number 
that represents the ratio of viscous dissipation to heat conduction with 
$c_p=7/2$ as the specific heat at constant pressure for a diatomic gas.

\FloatBarrier


\begin{figure}
\centering
 \includegraphics[scale=0.55]{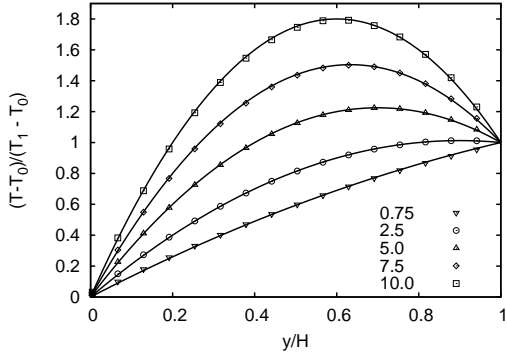}
\caption{Temperature profiles at steady state (symbols) compared to the 
analytical solution (lines) at varying Prandtl numbers. }
\label{fig:eckert}
\end{figure}

Simulations were performed for ${\rm Pr} = 0.75,\ 2.5,\ 5.0, \ 7.5$, and $10$ at
Eckert number fixed at unity on a domain with $128$ grid points.
The walls were maintained at temperatures 
$\theta_0+0.5\Delta \theta$ and $\theta_0-0.5\Delta \theta$ and plate velocity
$U_0$ is chosen corresponding to a Mach number of $0.1$. 
Figure \ref{fig:eckert} compares the 
temperature profiles obtained analytically and via simulations and they are
found to be in good agreement.

\section{Outlook}\label{sec:conclusion}
We have proposed a kinetic model for polyatomic gases 
with a tunable Prandtl number. This model is based on the ES--BGK model and recovers the compressible hydrodynamic equations of polyatomic gases as its macroscopic limit. It was shown that the transport coefficients of the model can be tuned for simulation of flows at different Prandtl numbers and specific heat ratios. This framework is general enough to deal with a more complex model of internal structures. We also demonstrated that the model respects the $H$ theorem. The simplicity of the model makes it suitable for LB and other numerical implementations.   
\appendix

\section{Evolution equations}\label{evolEqns}
We derive the evolution equations for kinetic energy, internal 
energy, pressure, stress, heat flux, translational, and rotational 
temperatures for the proposed model.
Using the momentum evolution equation Eq.\eqref{ConservedMomEq}, we obtain the 
evolution equation for $ \rho u_\alpha u_\beta$ as
\begin{equation}
\partial_t \left( \rho u_\alpha u_\beta \right)  + \partial_\gamma (\rho 
u_\alpha u_\beta u_\gamma) 
+ u_\alpha \partial_\gamma ( \rho \theta_T \delta_{\beta \gamma} 
+ \sigma_{\beta \gamma} ) 
+ u_\beta  \partial_\gamma  (\rho \theta_T \delta_{\alpha \gamma}
+ \sigma_{\alpha \gamma} ) = 0 .
\end{equation}

Evolution of kinetic energy obtained by taking the trace of the above equation
is 
\begin{equation}
 \partial_t \left(\frac{1}{2}\rho u^2\right)
+ \partial_\gamma \left( \frac{1}{2} \rho u^2 u_\gamma \right) 
+ u_\beta \partial_\gamma  (\rho \theta_T \delta_{\beta \gamma}
+ \sigma_{\beta \gamma} ) = 0.
\label{KE}
\end{equation}
Subtracting the above equation from the evolution equation of total energy
from Eq.\eqref{ConservedMomEq}, gives the evolution equation for internal energy
$e = (3+\delta)\rho \theta/2$ as
\begin{align}
\begin{split}
  \partial_t e 
 + \partial_\beta  \left( e u_\beta\right)
 + \partial_\beta q_\beta 
 + \sigma_{\beta \gamma} \partial_\gamma u_\beta
 +  \rho \theta_T  \partial_\gamma u_\gamma  = 0,
\end{split}
\label{internalEnergy}
\end{align} 
and the evolution equation of pressure $p = \rho \theta$ as
\begin{align}
\begin{split}
  \partial_t p
 + \partial_\beta  \left( p u_\beta\right)
 + \left(\frac{2}{3+\delta}\right) \left(\partial_\beta q_\beta 
 + \sigma_{\beta \gamma} \partial_\gamma u_\beta
 +  \rho \theta_T  \partial_\gamma u_\gamma \right) = 0.
\end{split}
\label{pressure}
\end{align}
Using  the pressure and continuity equation, the 
evolution equation for temperature $\theta$ is 
\begin{align}
\begin{split}
  \partial_t \theta
 + u_\beta \partial_\beta \theta
 + \left(\frac{2}{(3+\delta)\rho}\right) \left( \partial_\beta q_\beta 
 + \sigma_{\beta \gamma} \partial_\gamma u_\beta
 + \rho \theta_T  \partial_\gamma u_\gamma \right) = 0.
\end{split}
\label{temperature}
\end{align} 

From the evolution of kinetic energy (Eq.\eqref{KE}) and the translational 
temperature (Eq.\eqref{keES--BGKeq1}), the evolution for 
translational energy can be evaluated as
\begin{align}
\begin{split}
 \partial_t \left( \frac{3 \rho \theta_T}{2} \right) 
 +  \partial_{\beta} \left( \frac{3 \rho \theta_T}{2} u_\beta  \right)
 + \rho \theta_T \partial_\beta u_\beta 
 + \sigma_{\beta\gamma} \partial_\beta u_\gamma
 + \partial_\beta q^{T}_{\beta} = 
\frac{\rho}{\tau_1} \left(\frac{3}{2}\theta -  \frac{3}{2}\theta_T \right).
\end{split}
\label{translationalEnergy}
\end{align} 
Using the continuity equation, 
the evolution  for translational temperature $\theta_T$ is
\begin{align}
\begin{split}
 \partial_t \theta_T
 + u_\beta \partial_{\beta} \theta_T
 + \frac{2}{3} \theta_T \partial_\beta u_\beta 
 + \frac{2}{3 \rho} \sigma_{\beta\gamma} \partial_\beta u_\gamma
 + \frac{2}{3 \rho} \partial_\beta q^{T}_{\beta} = 
\frac{1}{\tau_1} \left(\theta - \theta_T \right).
\label{translationalTemp}
\end{split}
\end{align}

For the evolution of the stress tensor $\sigma_{\alpha \beta}$, we multiply the 
kinetic equation (Eq.\eqref{eq:kinEqBGK}) with $\xi_\alpha \xi_\beta$ and 
integrate over the velocity space to obtain

\begin{align}
\begin{split}
& \partial_t (\rho \theta_T \delta_{\alpha \beta}+ \sigma_{\alpha \beta})
+ \partial_\kappa Q^{T}_{\alpha \beta \kappa}
+ \partial_\kappa \left(u_\kappa  (\rho 
\theta_T \delta_{\alpha \beta} + \sigma_{\alpha \beta} )\right)
+  (\rho \theta_T \delta_{\kappa \beta} + 
\sigma_{\kappa \beta} ) \partial_\kappa u_\alpha  \\
& + (\rho \theta_T \delta_{\kappa \alpha} + 
\sigma_{\kappa \alpha}) \partial_\kappa u_\beta  = \frac{1}{\tau}  \left( b -1 
\right) \sigma_{\alpha \beta} +\frac{1}{\tau_1} \left( \rho \theta 
\delta_{\alpha \beta}  - \rho \theta_T \delta_{\alpha \beta} - \sigma_{\alpha 
\beta} \right).
\end{split}
\label{stressMajorEqn}
\end{align}

Thereafter, multiplying Eq.\eqref{translationalEnergy} with $\delta_{\alpha 
\beta}$ and subtracting from the above 
equation one obtains evolution of stress tensor as
\begin{equation}
 \partial_t (\sigma_{\alpha \beta})
+ \partial_\gamma \left(u_\gamma  \sigma_{\alpha \beta} \right)
+ \partial_\gamma \overline{Q^{T}_{\alpha \beta \gamma}} 
+ \frac{4}{5} \overline{\partial_\beta  q^{T}_\alpha}
+ 2 \rho \theta_T \overline{\partial_\beta u_\alpha}
+ 2 \overline{\partial_\gamma u_\alpha \sigma_{\gamma \beta}}
= \left(\frac{1}{\tau}\left(b-1\right) - \frac{1}{\tau_1} \right) 
\sigma_{\alpha \beta}.
\label{stressEvolution}
\end{equation}

Similarly, the evolution equation for translational heat flux is obtained by 
multiplying the kinetic equation (Eq.\eqref{eq:kinEqBGK}) with $\xi^2 \xi_\alpha$ to obtain
\begin{align}
\begin{split}
& \partial_t q^{T}_\alpha 
+ \partial_\beta \left(u_\beta q^{T}_\alpha \right)
+ \overline{Q^{T}_{\alpha \beta \gamma}} \partial_\beta u_\gamma
+ \frac{1}{2} \partial_\beta R_{\alpha \beta} 
+ \frac{7}{5}  q^{T}_\beta 
\partial_\beta u_\alpha
    + \frac{2}{5} q^{T}_{\alpha} \partial_\eta u_\eta 
    + \frac{2}{5} q^{T}_{\beta} \partial_\alpha u_\beta \\
&- \frac{5}{2} \theta_T \partial_\alpha (\rho \theta_T)  
  - \frac{\sigma_{\alpha \beta}}{\rho} \partial_\beta(\rho \theta_T) 
 - \frac{5}{2} \theta_T  \partial_\kappa \sigma_{\kappa \alpha} 
 - \frac{\sigma_{\alpha \beta}}{\rho} \partial_\kappa \sigma_{\kappa \beta}  
= -\left(\frac{1}{\tau} + \frac{1}{\tau_1} \right)  q^{T}_\alpha.
\end{split}
\label{translationalHeatFluxEvolutionEquation}
\end{align}

\bibliographystyle{jfm}
\bibliography{PrandtlCorrection_diatomic.bib}

\end{document}